\documentclass[iop]{emulateapj}
\pdfoutput=1
\usepackage{graphicx}
\usepackage{pdfpages}  %Compile with pdflatex, not latex. Convert eps figs with pstopdf
\newcommand{\myemail}{tom.greene@nasa.gov}
\slugcomment{Accepted by ApJ 2015 November}

\shorttitle{JWST exoplanet characterization}
\begin{document}

\title{Characterizing transiting exoplanet atmospheres with JWST}

\author{Thomas P. Greene\altaffilmark{1}}
\affil{NASA Ames Research Center, Space Science and Astrobiology Division, 
M.S. 245-6; Moffett Field, CA 94035}
%\email{tom.greene@nasa.gov}
\author{Michael R. Line\altaffilmark{2,3,4,5}}
\author{Cezar Montero}
\author{Jonathan J. Fortney}
\affil{Department of Astronomy and Astrophysics, University of California, Santa Cruz, CA 95064}
\author{Jacob Lustig-Yeager}
\affil{Department of Astronomy, Box 351580, University of Washington, Seattle, WA 98195}
\author{Kyle Luther}
\affil{Department of Physics, University of California, 366 LeConte Hall MC 7300, Berkeley, CA 94720}
\altaffiltext{1}{Correspondence to be directed to \myemail}
\altaffiltext{2}{Hubble Postdoctoral Fellow}
\altaffiltext{3}{NASA Ames Research Center, Moffett Field, CA}
\altaffiltext{4}{Bay Area Environmental Research Institute, Petaluma, CA}
\altaffiltext{5}{School of Earth and Space Exploration, Arizona State University, Tempe, AZ}

\begin{abstract}

We explore how well James Webb Space Telescope (JWST) spectra
will likely constrain bulk atmospheric properties of transiting
exoplanets. We start by modeling the atmospheres of archetypal hot
Jupiter, warm Neptune, warm sub-Neptune, and cool super-Earth planets
with clear, cloudy, or high mean molecular weight atmospheres. Next we
simulate the $\lambda = 1-11$ $\mu$m transmission and emission spectra
of these systems for several JWST instrument modes for single transit
and eclipse events. We then perform retrievals to determine how well
temperatures and molecular mixing ratios (CH$_4$, CO, CO$_2$, H$_2$O,
NH$_3$) can be constrained. We find that $\lambda = 1-2.5$ $\mu$m
transmission spectra will often constrain the major molecular
constituents of clear solar composition atmospheres well. Cloudy or
high mean molecular weight atmospheres will often require full $1-11$
$\mu$m spectra for good constraints, and emission data may be more
useful in cases of sufficiently high $F_p$ and high $F_p/F_*$. Strong
temperature inversions in the solar composition hot Jupiter atmosphere
should be detectable with $1-2.5+$ $\mu$m emission spectra, and
$1-5+$ $\mu$m emission spectra will constrain the temperature-pressure
profiles of warm planets. Transmission spectra over $1-5+$ $\mu$m will
constrain [Fe/H] values to better than 0.5 dex for the clear
atmospheres of the hot and warm planets studied. Carbon-to-oxygen
ratios can be constrained to better than a factor of 2 in some
systems. We expect that these results will provide useful predictions
of the scientific value of single event JWST spectra until its on-orbit
performance is known.

\end{abstract}


\section{Introduction} \label{sec:Intro}

There are now well over a thousand confirmed exoplanets, ranging from
hot to cold and large to small worlds. Characterizing the atmospheres
of a diversity of planets is critical to understanding their bulk
compositions, formation (and any migration), energy balance, and
atmospheric processes \citep[e.g., see][]{C15, MKF14, BO10, HS15,
SD10, B14}.

Much of the atmospheric characterization work has come from the
acquisition and interpretation of transmission and emission
spectroscopy of transiting planets by the Hubble and Spitzer space
telescopes. The transmission or emission spectra of Hot Jupiters (e.g.,
HD 189733b, WASP-43b), warm Neptunes (e.g., GJ 436b, HAT-P-11b), and
warm sub-Neptunes (e.g., GJ 1214b) are all being studied extensively
with HST, Spitzer, and other facilities. The near-IR spectra of cool or
warm super-Earths like K2-3b/c/d will soon be observed with HST, and
they will likely be prime candidates for JWST spectroscopy.
Observations to date have produced a variety of important discoveries
such as the detection of H$_2$O absorption, now clearly seen in a
variety of planets using a variety of instruments. HST NICMOS and HST
WFC3 G141 find strong water absorption in about a dozen hot Jupiters
\citep[e.g.,][]{KBD14b, KLB15, DWM13, CMB12, SVT09} and a warm
Neptune-sized planet \citep{FDB14}. Atmospheric retrieval techniques
have been applied to the data in order to determine the abundances (or
upper limits) of molecules such CO, CO$_2$, and CH$_4$ in addition to
H$_2$O in $\sim 10$ exoplanet atmospheres \citep[e.g.,][]{MS09, MHS11,
LFI12, LZV12, BAI13a, BAI13, BS12, BS13, LWZ13, LKW14, B15, WTR15}.
Molecular abundance determinations have been used to constrain
atmospheric C/O ratios \citep[e.g.,][]{B15, LKW14, MHS11} which can
potentially help diagnose where a planet formed relative to the H$_2$O
and CO ice lines in its protoplanetary disk \citep{OMB11}. Atmospheric
metallicity determinations relative to the host star have also been
used to infer formation possibly via a core accretion
\citep[e.g.,][]{KBD14b}. Temperature inversions have been suggested to
explain the emission spectra of a number of hot Jupiters \citep[e.g.,
see][and references therein]{FLM08, KHI10}, indicating the presence of
visible or UV stratospheric absorbers. However, recent work by
\citet{LKW14} indicates no strong statistical evidence for inversions
in a sample of 9 observed planets, but the very hot Jupiters HAT-P-7b
\citep{CBC10} and WASP-33b \citep{HMM15} do appear to have temperature
inversions at the present time \citep[see also][]{C15}. Determining the
frequency of inversions over a variety of bulk planetary properties is
important for understanding chemical processes (e.g., impact of C/O on
high altitude absorbers) and the overall energy balance in these
planets' atmospheres.

Despite these advances, there are still considerable uncertainties in
the compositions, temperatures, and origins of exoplanet atmospheres.
Numerous early HST and Spitzer detections of molecular features and
temperature inversions have been called into question or disproven with
subsequent higher precision observations, more sophisticated data
analysis, and powerful modern retrieval techniques for molecular
abundances and temperature-pressure (hereafter T-P) profiles
\citep[e.g.,][]{B15, SBd15, LKW14, D-LSB14, GPA11}. Clearly, we
need to better determine the compositions of exoplanet
atmospheres, what planets have stratospheric temperature inversions
under what conditions, how closely planet elemental abundances match
their host stars, and where planets formed in their disks. Addressing
or resolving these specific questions would significantly advance our
understanding of exoplanet atmospheres:

\begin{itemize}
	
\item[] Do any highly insolated planets have hot stratospheres, and
what absorbers are causing these temperature inversions if they exist?

\item[] What are the nature of super-Earth (1.5 -- 2 R$_\earth$)
planet atmospheres; are they mostly H and He or are they dominated by
high mean molecular weight species like Earth and Venus?

\item[] How do clouds inhibit our ability to infer molecular abundances?

\item[] What are the C-to-O ratios in planetary atmospheres, how does
this compare to their host stars, and what does this imply about where
planets formed in protoplanetary disks?

\item[] How to the metal abundances of planets compare to their host
stars, and does this vary by planet mass as it does for giants in our solar system?

\item[] How far from chemical equilibrium can exoplanet atmospheres be driven, and what causes this??

\end{itemize}

High quality JWST observations may characterize transiting exoplanet
atmospheres well enough to address these questions significantly in the
near future (launch is currently scheduled for 2018 October). JWST's
large aperture (6.5-m), numerous spectroscopic modes over $\lambda = 0.6
- 28$ $\mu$m, good thermal stability, and applications of lessons
learned from other observatories will ensure that it collects the
highest quality exoplanet transmission and emission spectra. Numerous
studies are providing assessments of how well JWST is expected to
characterize exoplanet atmospheres. \citet{BBK14} present many details
of JWST's instruments and recommend appropriate modes for observing
transiting exoplanets. \citet{CGA15} report that JWST is expected to
characterize dozens of giant planets over its mission lifetime, but
observations of cool, small planet (``temperate terrestrial'')
atmospheres may require $\sim$100 days each. \citet{BKL15} find that
JWST NIRSpec should be able to measure the $\lambda = 1 - 5$ $\mu$m
transmission spectra of nearby (3 -- 50 pc), cool ($400 - 1000$ K),
low-mass ($1 - 10 M_\earth$) planets orbiting mid-M dwarfs with moderate
signal-to-noise ratios (SNRs) after summing 25 transits. \citet{BAI15}
performed JWST data simulations and atmospheric retrievals to
investigate the impacts of starspots and other systematic errors that
shift different spectral wavelength ranges that are not obtained
simultaneously.

These studies have greatly improved our understanding of the promises
and potential limits of JWST data. However, we still do not understand
how well JWST will be able to quantitatively characterize different
types of planets in the above ways. We also need to understand better
what observing modes will be most useful for addressing specific
questions. This is an important assessment because most transiting
planets with bright host stars ($J \lesssim 11$ mag) will need to be
observed 4 separate times to collect their entire $\lambda = 0.7 - 12+$
$\mu$m spectra \citep[e.g., see][]{BBK14}, and JWST time will be
extremely precious. We perform and discuss such quantitative
assessments in this contribution.

We investigate how well temperature and molecular volume mixing ratio
constraints from JWST observations can address these big picture
questions by analyzing simulated observations of a diverse range of
planet types. We start by presenting a diverse set of planetary systems
that span what we think are the typical planet types in
\S\ref{sec:Systems}. We describe the atmospheric models and the
retrieval technique in \S\ref{sec:Models}. Next we describe the
instrument signal and noise models we use to create simulated JWST
spectra in \S\ref{sec:Sims}. The retrieval results including the
molecular and temperature constraints are presented in
\S\ref{sec:Results}. We apply these results to assess planet
carbon-to-oxygen ratios, metallicities, and probes of disequilibribum
chemistry in \S\ref{sec:Discussion} and the big picture questions these
can address. Finally, we summarize our conclusions in
\S\ref{sec:Summary}.

\section{Simulated Planets}\label{sec:Systems}

We assemble a set of 4 fiducial planetary systems to assess how well
JWST will be able to characterize exoplanet atmospheres in the early
years of its mission. These planets range from hot ($T_{\rm eq} =
1500$ K) to cool ($T_{\rm eq} = 500$ K) and large (1.36 $R_{\rm J}$) to
small (0.19 $R_{\rm J}$). This combination of sizes and temperatures
maps well onto the now established planet archetypes of hot Jupiters,
warm Neptunes, warm sub-Neptunes, and cool super-Earths. We select the
physical parameters of a well known system from each of these
categories to assemble our set of 4 systems to model. We model their
atmospheres as either clear with solar elemental abundances, cloudy
with solar elemental abundances, and for the smaller planets clear
with enhanced elemental abundances so as to result in high mean
molecular weight (HMMW) atmospheres.

Constant-with-altitude molecular abundances were generated assuming
broad consistency with thermochemical equilibrium given the effective
temperature of the planet and the elemental abundances (computed with
the Chemical Equilibrium with Applications code \citep{GM94, VM11,
LLY10, MVF11}\footnote{We note that thermochemical equilibrium generally
does not produce constant-with-altitude mixing ratio profiles. However,
for the dominant carbon-bearing species and H$_2$O, constant with
altitude generally does occur in equilibrium. For the less dominant
species we choose a representative value along a non-uniform vertical
profile.}, and the HMMW atmospheres were either $1000 \times$ solar
metallicity or pure H$_{2}$O. The $1000 \times$ solar compositions
correspond to mean molecular weight $\mu$=15 for the warm Neptune and
$\mu$ = 16.8 for the warm sub-Neptune atmospheres (see \citet{FMN13}).
It is not clear whether $1000 \times$ solar metallicity is possible
since accreted planetesimals include hydrogen if they are icy
\citep{FMN13}, preventing this level of atmospheric metallicity.
Nevertheless, we believe that this is a useful bounding case. The 100\%
H$_2$O cool super-Earth atmosphere has a molecular weight of $\mu$ = 18
and corresponds to an [Fe/H] = 2.77 ($588 \times$ solar
metallicity)\footnote{It is impossible to generate a pure H$_2$O
atmosphere thermochemically by scaling the solar elemental abundances.
The metallicity value here is determined by taking 0.5 (the ratio of
metals, O, to hydrogen (2H)) and dividing that by the solar metal
fraction value of 8.5$\times$10$^{-4}$}. The set of planet types, system parameters, and atmospheric molecular mixing ratios are given in Tables \ref{tbl-1}, \ref{tbl-2}, and \ref{tbl-mol}. Actual small, cool planets may not be in chemical equilibrium, and we discuss how well we can detect disequilibrium chemistry in \S\ref{sec:Disequilibrium}.

\begin{deluxetable*}{lllll}
\tabletypesize{\scriptsize}
%\rotate
\tablecaption{Planetary Systems Modeled\label{tbl-1}}
\tablewidth{0pt}
\tablehead{
\colhead{Planet Type} & \colhead{System Parameters} & \colhead{Composition} & 
\colhead{Clouds} & \colhead{Geometry}
}
\startdata
Hot Jupiter			& HD 209458b & $1\times$ Solar    & Clear	&  Trans, Emis\\
             		&			&             & 1 mbar	&  Trans\\
\\
Warm Neptune 		& GJ 436b	& $1\times$ Solar    & Clear	&  Trans, Emis\\
			 		&			&			  & 1 mbar	&  Trans\\
			 		&			& $1000\times$ Solar & Clear	&  Trans, Emis\\
\\			
Warm Sub-Neptune 	& GJ 1214b	& $1\times$ Solar    & Clear	&  Trans, Emis\\
					&			&             & 1 mbar	&  Trans\\
					&			& $1000\times$ Solar & Clear	&  Trans, Emis\\
\\
Cool Super-Earth	& K2-3b		& $1\times$ Solar	  & Clear	&  Trans, Emis\\
					&			&             & 1 mbar	&  Trans\\
					&			& 100\% H$_{2}$O & Clear	&  Trans, Emis
\enddata
\end{deluxetable*}

\begin{deluxetable*}{llrrrrrrrr}
\tabletypesize{\scriptsize}
%\rotate
\tablecaption{Fiducial Planetary System Parameters\label{tbl-2}}
\tablewidth{0pt}
\tablehead{
\colhead{Planet Type} & \colhead{System Parameters} &
\colhead{$T_{*}$ (K)} & \colhead{$R_{*}$ ($R_\odot$)} & 
\colhead{$K$ (mag)} & \colhead{$T_{\rm eq}$\tablenotemark{a} (K)} & 
\colhead{$M_{\rm p}$ ($M_\oplus$)} & \colhead{$R_{\rm p}$ ($R_\oplus$)} &
\colhead{H\tablenotemark{b} (km)} & \colhead{$T_{14}$ (s)}
}
\startdata
Hot Jupiter			& HD 209458b & 6065 & 1.155 & 6.3 & 1500 & 220 & 15 & 560 & 11,000 \\
Warm Neptune 		& GJ 436b	& 3350 & 0.464 & 6.1 & 700 & 23 & 4.2 & 190 & 2740 \\
Warm Sub-Neptune 	& GJ 1214b	& 3030 & 0.211 & 8.8 & 600 & 6.5 & 2.7 & 230 & 3160 \\
Cool Super-Earth	& K2-3b		& 3900 & 0.561 & 8.6 & 500 & 5.3\tablenotemark{c} & 2.1 & 150\tablenotemark{c} & 9190 
\enddata
\tablecomments{Tabulated system values were taken from the exoplanets.org  compilation \citep{HWW14} and \citet{CPS15}.}
\tablenotetext{a}{Equilibrium temperature $T_{\rm eq}$ was computed from the listed system values assuming albedo = 0 and energy re-distribution over 4$\pi$ str.}
\tablenotetext{b}{The planetary atmosphere scale height H = $k T_{\rm 
eq} / (\mu m_{H} g)$ for the clear solar atmosphere of each planet 
($\mu = 2.3$) is provided as a convenience for scaling to other systems.}
\tablenotetext{c}{The mass of this planet has been recently measured to be $8.4 \pm 2.1 M_\oplus$ \citep{AAB15}, somewhat higher than the tabulated value we used in our investigation. This increased mass
would decrease the scale height, decrease the SNR of transmission spectral features, and worsen the derived abundance precisions by roughly 40\%.}
\end{deluxetable*}

\begin{deluxetable*}{llrrrrrr}
\tabletypesize{\scriptsize}
\tablecaption{Planetary Atmosphere Molecular Volume Mixing Ratios\label{tbl-mol}}
\tablewidth{0pt}
\tablehead{
\colhead{Planet Type} & \colhead{Composition} & 
\colhead{H$_2$O} & \colhead{CH$_4$} & 
\colhead{CO} & \colhead{CO$_2$} & 
\colhead{NH$_3$} & \colhead{N$_2$}
}
\startdata
Hot Jupiter     & $1\times$ Solar & 4.27E-4 & 1.00E-9 & 4.27E-4 & 1.26E-7 & 3.16E-10 & 5.75E-5 \\
Warm Neptune    & $1\times$ Solar & 7.24E-4 & 4.27E-4 & 1.00E-9 & 3.16E-11 & 3.16E-5 & 2.51E-5 \\
Warm Neptune & $1000\times$ Solar & 2.51E-1 & 1.00E-1 & 1.00E-2 & 1.45E-1 & 5.01E-5 & 4.47E-2 \\
Warm Sub-Neptune & Solar & 7.24E-4 & 4.26E-4 & 1.00E-9 & 3.16E-11 & 3.16E-5 & 2.51E-5 \\
Warm Sub-Neptune & $1000\times$ Solar & 3.98E-1 & 1.26E-1 & 1.00E-3 & 1.26E-1 & 5.01E-5 &  5.01E-2 \\
Cool Super-Earth & $1\times$ Solar & 7.24E-4 & 4.27E-4 & 1.00E-11 & 1.00E-11 & 1.00E-4 & 1.58E-5 \\
Cool Super-Earth & 100\% H$_2$O & 1.00E00 &  0 & 0 & 0 & 0 & 0
\enddata
\tablecomments{All mixing ratios are assumed constant with altitude and are thermochemical equilibrium values for the planet temperatures and compositions given in Tables \ref{tbl-1} and \ref{tbl-2}\, except the Hot Jupiter terminator temperature of 1200 K \citep{MVF11} was used instead of its Table \ref{tbl-2} $T_{\rm eq}$ value.
}
\end{deluxetable*}

\section{Modeling and Retrieval Approach} \label{sec:Models} 

The emission spectra are computed using the forward model described in
\citet{LWZ13} and subsequent upgrades described in \citet{D-LSB14} and
\citet{SBM14}. The transmission spectra are computed with the forward
model described in \citet{LKD13}, \citet{SLD14}, and \citet{KBD14b}.
We then use a derivative of the CHIMERA\footnote{Instead of using the
Differential Evolution Markov chain Monte Carlo as in \citet{LKD13,
LKW14}, we use the EMCEE routine of \citet{F-MHM14}} retrieval suite
\citep{LWZ13,LKD13} to determine the degree of constraints on
temperatures and abundances from simulated observations of the
transmission and emission spectra (see \S\ref{sec:Sims}). The molecular opacities described in \S2.3 of \citet{LTB15} were used for computing all forward and retrieval model spectra.

While atmospheres are undoubtably complicated, we choose a relatively
simple 1-D parameterization. Since we are using the same forward model
to generate the synthetic spectra as the retrieval, this should not be a
problem for assessing the impact of JWST data quality. This will allow
us to strictly explore the role that the JWST instrumental noise
properties and spectral coverage have on setting the atmospheric
constraints. We perform a preliminary exploration of retrieval
assumptions and priors and their inherent biases on the retrieved
results later in this work (see \S\ref{sec:Parameterization}), and a more thorough analysis will be the subject of a forthcoming paper.

\begin{figure*}
\begin{center}
\includegraphics[width=0.46\textwidth, angle=0]{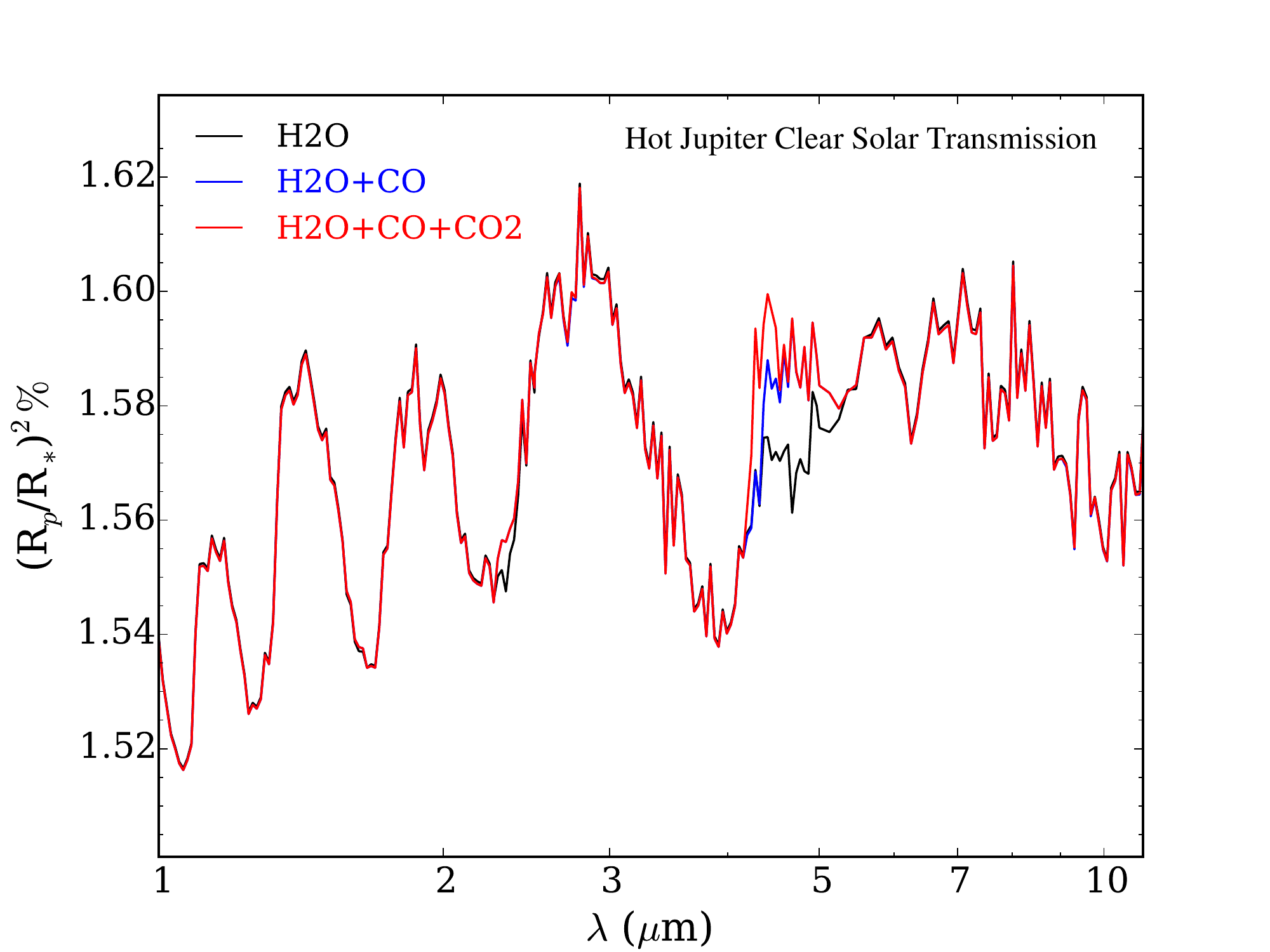}
%\includegraphics[width=1.67in, angle=0]{spectra_sensitivity_hot_Jupiter_transmission.pdf}
%\hspace{0.01 in}%
\includegraphics[width=0.46\textwidth, angle=0]{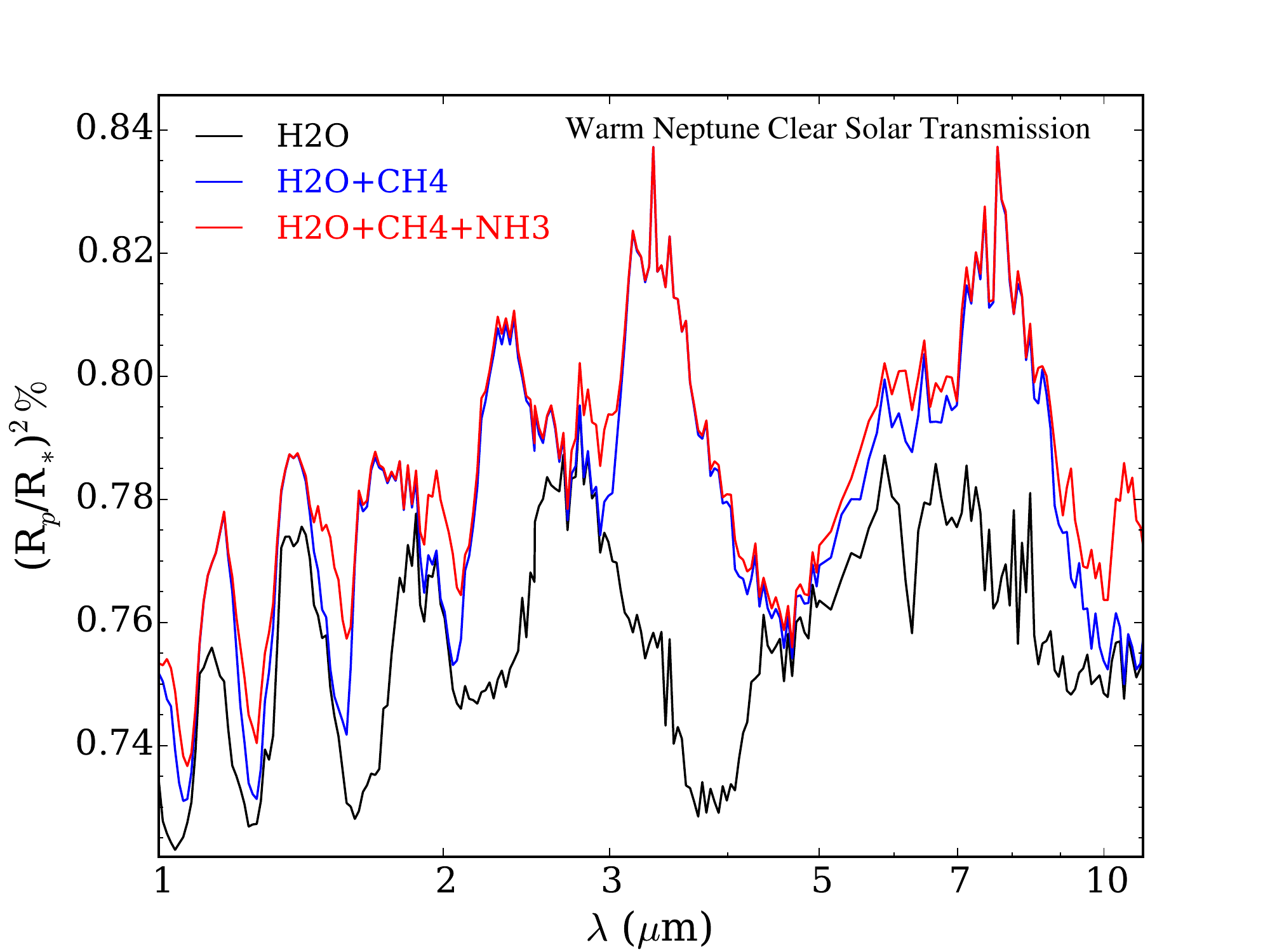}
\end{center}
\begin{center}
%\hspace{0.1 in}%
\includegraphics[width=0.46\textwidth,]{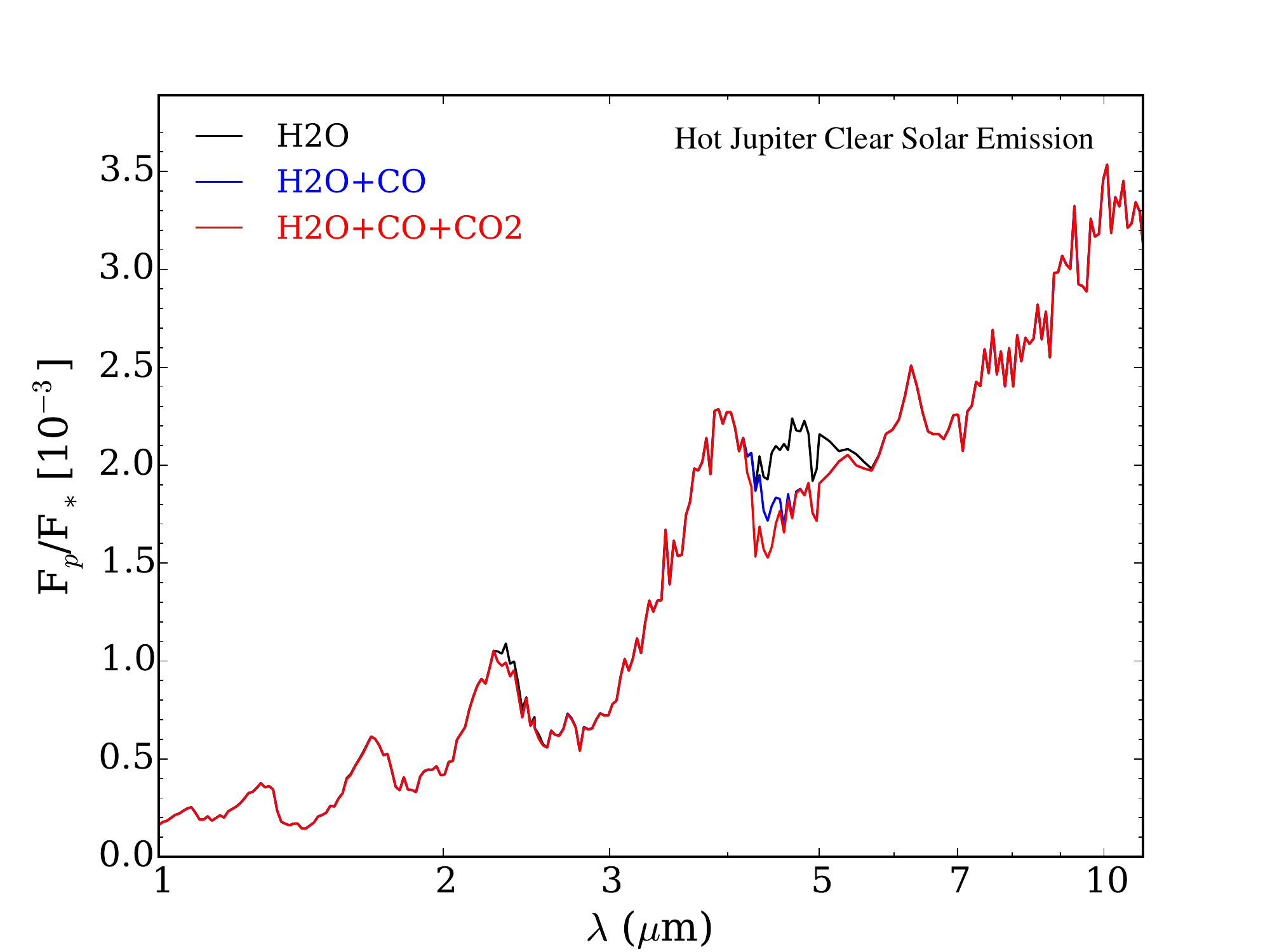}
\includegraphics[width=0.46\textwidth,]{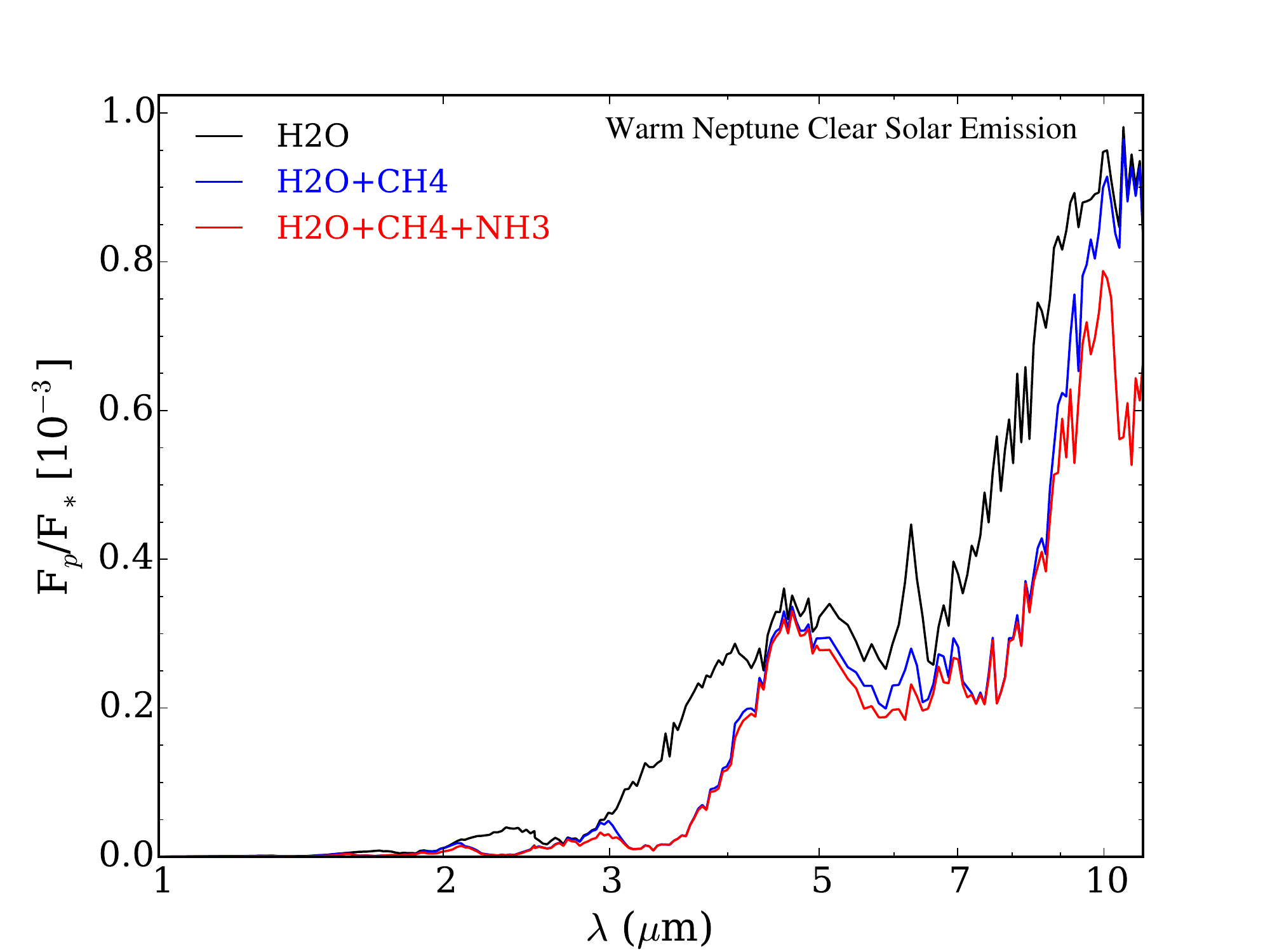}
\end{center}   
     \caption{ \label{fig:gas_perturbations_spec_fig} 
Left: Magnitude of absorption for specific gases for the clear solar composition hot Jupiter transmission (upper left) and emission (lower left) models.  
Right: Wavelengths and amplitudes of gaseous absorptions for the clear solar composition warm Neptune transmission (upper right) and emission (lower right) models. Each panel shows the impact of the dominant molecules in each model.
  }
\end{figure*}

Because of the different viewing geometries, the atmospheres used in
transmission are parameterized differently than the atmospheres used to
generate the emission spectra. However, both use the same set of
uniform-with-altitude molecular abundances. These include H$_2$O,
CH$_4$, CO, CO$_2$, NH$_3$, and N$_2$ (except in emission as it has no
consequence on the emission spectra). Any remaining gas is assumed to
be a mixture of solar composition H$_2$/He. N$_2$ is the dominant
nitrogen bearing species at high temperatures, but it has no
spectroscopic features unless in high concentrations \citep{SRM15}. We
include it here as a trace gas simply to contribute to the mean
molecular weight of atmospheres in transmission spectra. The
fiducial abundances are again chosen to be broadly consistent with
thermochemical equilibrium at the given scale height temperature. For
all but the hot-Jupiter scenario we deplete 2-oxygen atoms per every
magnesium to emulate oxygen loss due to enstatite (Mg$_2$SiO$_4$)
condensation. The impact of various absorption features for each of
these species in the clear solar composition hot Jupiter and warm
Neptune transmission and emission model spectra in
Figure~\ref{fig:gas_perturbations_spec_fig}.

The transmission spectra are generated with 11 free parameters, 6 of
which are the volume mixing ratios of the molecular gasses noted above.
The remaining 5 parameters are as follows: First, an effective scale
height temperature ($T$). This will directly impact the amplitude of
the spectral features. For simplicity, we assume isothermal atmospheres
for the transmission spectra at the approximate equilibrium temperature
(without truly knowing the albedo or redistribution) of the planet in
question. \citet{BAI13a,BAI15} demonstrated that there may be potential
biases in real atmospheres if isothermal atmospheres are assumed, but
that it will be incredibly difficult to actually retrieve detailed
temperature $\it {profile}$ information from transmission spectra.
Second, we use a scaling to the fiducial 10 bar planet radius ($xR_p$).
This parameter manifests itself as a DC offset in the spectra as well
as a minor impact on the amplitude of the features. Third, we include
an opaque gray cloud parameterized with a cloud-top pressure ($P_{c}$)
which we set to 1 mbar. The atmospheric transmittance at
atmospheric levels deeper than the cloud top is set to zero. Several
measured flat exoplanet transmission spectra have been reasonably well
described with such a parameterization \citep{LKD13, KBD14a, KDK14}. We
also model hazes in an ad-hoc fashion using the following approximation
\citep[e.g.,][]{DVD08}:

\begin{equation}
 \sigma=\sigma_{0}(\frac{\lambda}{\lambda_0})^{-\beta}. 
\end{equation}

\begin{figure*}
\begin{center}
\includegraphics[width=1.0\textwidth, angle=0]{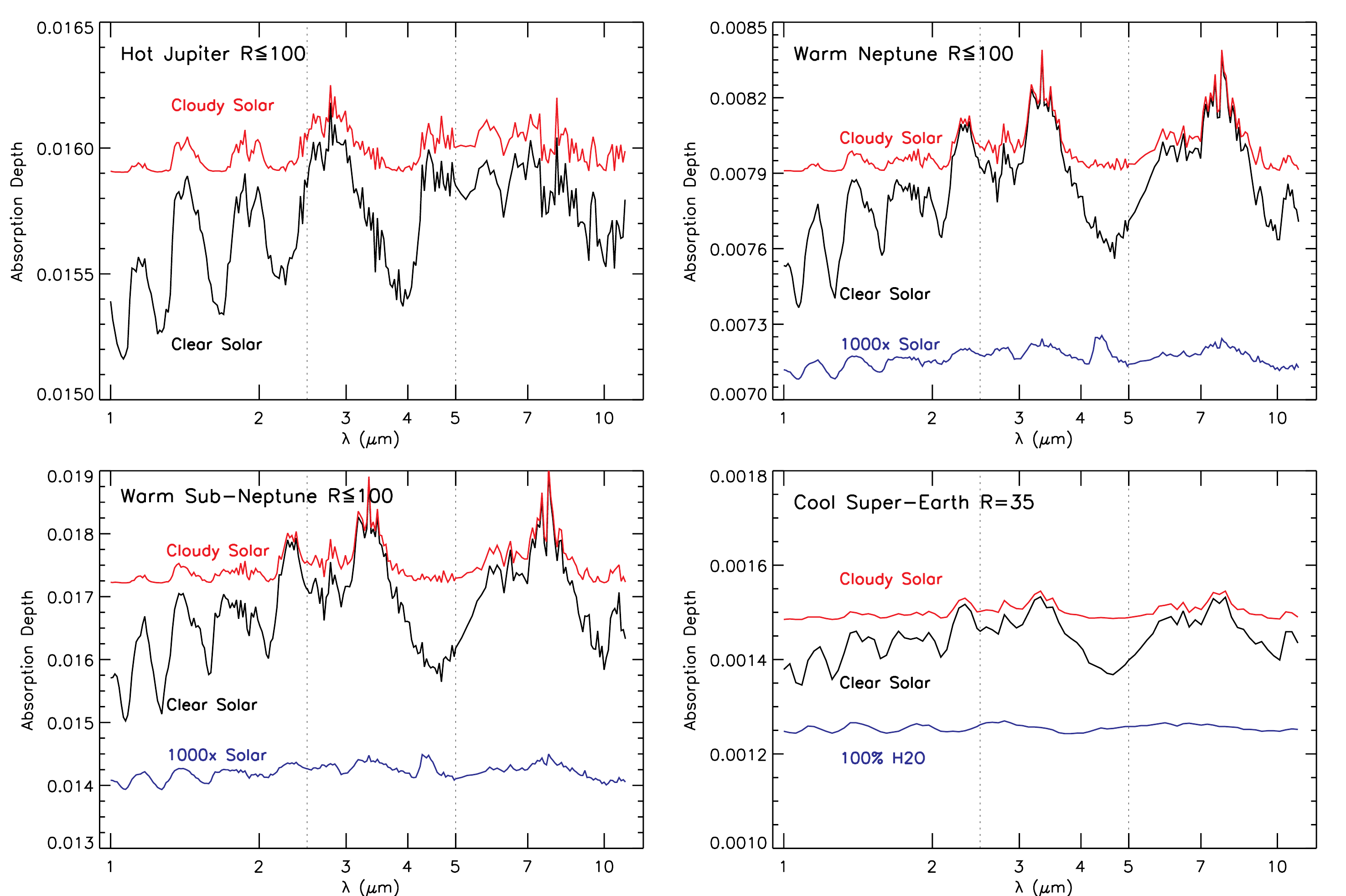}
\end{center}
     \caption{ \label{fig:forward_trans_spec_fig} 
Forward model transmission spectra of all planets in Table \ref{tbl-1}
where Absorption Depth = $(R_{p,\lambda}/R_{*})^2$. Spectra have been binned to resolution $R \leq 100$ (hot Jupiter, warm Neptune, warm
sub-Neptune; see \S\ref{sec:final_spectra}) or $R = 35$ (cool super-Earth) to match that shown for the simulated JWST data. Dashed lines show the wavelength range boundaries of the chosen NIRISS, NIRCam, and MIRI instrument modes. The vertical offsets in the cloudy spectra are due to increased opacity from opaque clouds, increasing the effective planetary radii \citep[e.g., see][]{B01}. The smaller
absorption depths of the high mean molecular weight atmospheres are
due to their smaller (than solar composition) scale heights. 
} 
\end{figure*}

$\sigma_{0}$ is the magnitude of the haze cross-section relative to
H$_2$ Rayleigh scattering at 0.4 $\mu$m, and $\beta$ is the wavelength
power law index that describes the slope. While the $\lambda \geq 1$
$\mu$m wavelength range explored in this study will not provide
constraints on these parameters, we include them because their
degeneracy with the absorbers at near-IR wavelengths could potentially
influence the retrieved abundances \citep[e.g.,][]{B15, KLB15}. Figure
\ref{fig:forward_trans_spec_fig} shows the spectra for all planets
modeled in transmission (see also Table \ref{tbl-1}).

\begin{figure*}
\begin{center}
\includegraphics[width=1.0\textwidth, angle=0]{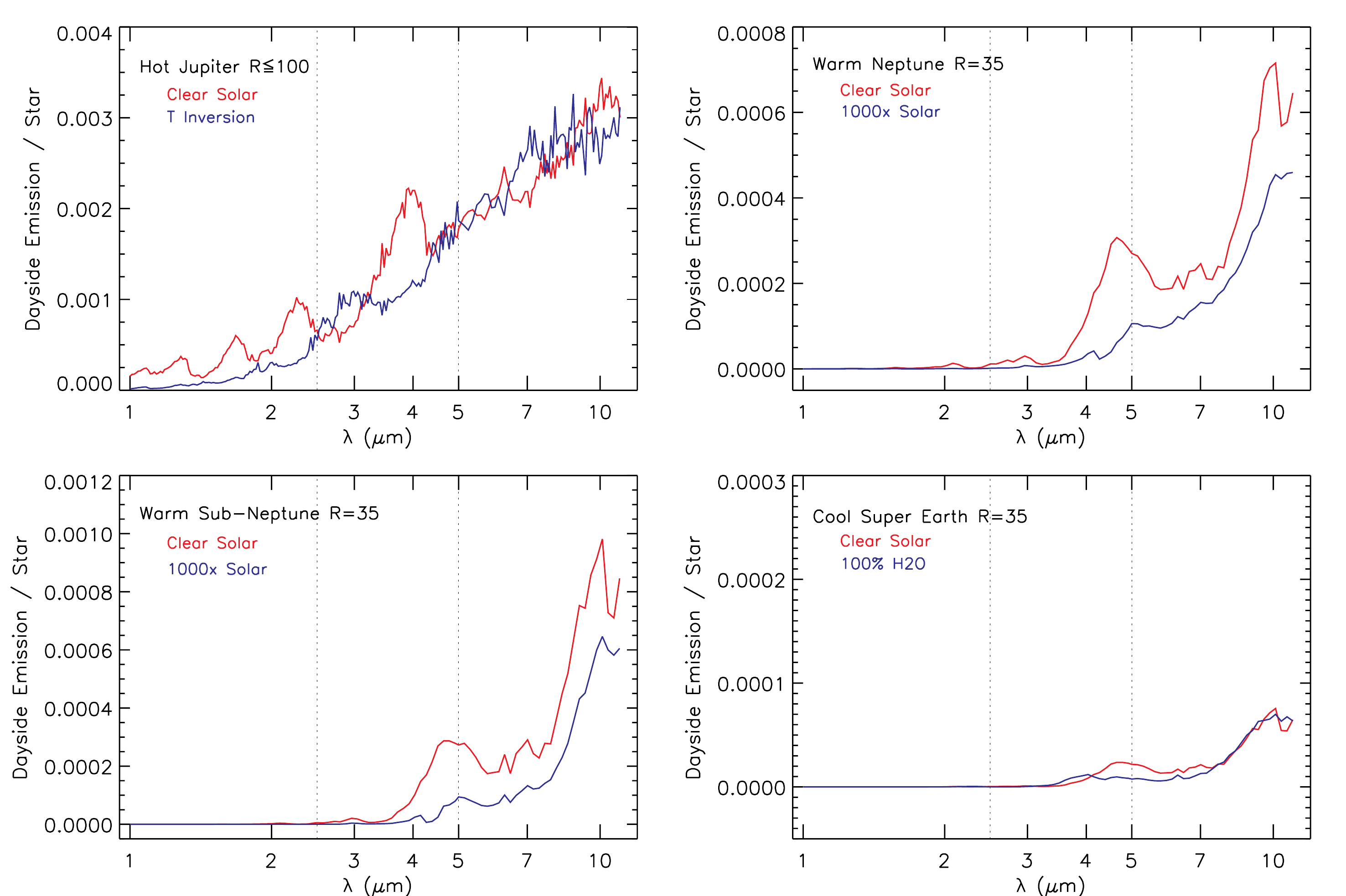}
\end{center}
     \caption{ \label{fig:forward_emis_spec_fig} 
Forward model emission spectra of all planets in Table \ref{tbl-1}.     
Spectra have been binned to resolution $R \leq 100$ (hot Jupiter, warm Neptune, warm sub-Neptune; see \S\ref{sec:final_spectra}) or $R = 35$ (cool super-Earth) to match that shown for the simulated JWST data. Dashed lines show the wavelength range boundaries of the chosen NIRISS, NIRCam, and MIRI instrument modes.
  }
\end{figure*}

The emission spectrum model requires 10 free parameters (5 of which are
the molecular absorbers). We assume a thermally and chemically
homogenous dayside, represented by a single 1D T-P profile. Because
thermal emission spectra contain temperature ``profile" information, we
use a 5 parameter double-gray analytic formula \citep[][and references
therein]{LWZ13}. We assume cloud-free emission spectra. Although clouds
can dominate transmission spectra, they may have a lessened impact on
emission spectra due to their reduced effective optical depths
\citep{F05}. The specific impact of clouds on emission spectra depends
on where (at what pressure) the thermal emission arises within the
atmosphere, where clouds form (i.e., where their material condensation
curves cross atmospheric T-P profiles), and the absorption and
scattering properties of all clouds within the pressure region probed
by the emission. This is complex enough to warrant its own
investigation, and we do not explore the impact of clouds on emission
spectra in this work. HST WFC3 dayside spectra of several hot Jupiters
are consistent with cloud-free atmospheres \citep{MCD14, SBM14, KSF15},
so we do know that clouds do not suppress spectral features in emission
spectra of all planets. Figure \ref{fig:forward_emis_spec_fig} shows
the spectra of all planets modeled in emission (see also Table
\ref{tbl-1}).

More complicated atmospheric processes such as vertical mixing,
atmospheric dynamics, photochemistry, ion chemistry (both photochemical
and thermal), unknown cloud processes, and perhaps many other known or
unknown processes (all active areas of research) could make our chosen
parameter values less than realistic. Certainly any of these processes
could be modeled with a nearly infinite number of unknown parameters.
However, we are only investigating the impacts of the spectral
(wavelengths and resolutions) and noise properties of JWST
instruments on retrieving information from their observations.
Therefore these assumptions are perfectly appropriate for this task of
assessing the uncertainties of retrievals and any systematic deviations
from the adopted forward models. We will use these parameters and their
constraints to evaluate the expected JWST performance.

We assume a uniform or uniform-in-log priors\footnote{We have also
experimented with the centered-log transform \citep{BS12} for both solar
and high mean molecular weight atmospheres and found no significant
qualitative differences} for the gases and other parameters, as done in
\citet{LFM14}, \citet{SBM14}, and \citet{KBD14b}. The retrievals are
initialized at the true parameter vector to minimize burn-in and to
speed convergence.

%%%%%%%%%%%%% figure %%%%%%%%%%%%%%%%%%%%
%%%%%%%%%%%% figure %%%%%%%%%%%%%%%%%%%%

%%%%%%%%%%%%% figure %%%%%%%%%%%%%%%%%%%%
 %%%%%%%%%%%% figure %%%%%%%%%%%%%%%%%%%%

%%%%%%%%%%%%% figure %%%%%%%%%%%%%%%%%%%%
%%%%%%%%%%%% figure %%%%%%%%%%%%%%%%%%%%

\section{Simulated JWST Spectra} \label{sec:Sims}

We have chosen to simulate the $\lambda = 1 - 11$ $\mu$m transmission
and emission spectra of the modeled systems. This region of the spectrum
shows the dominant absorption features for the major carbon-, oxygen-,
and nitrogen-bearing species. We do not focus on constraining alkali
metals and metal oxides/hydrides in this investigation. These species
(for hot planets), along with molecular Rayleigh scattering and possible
hazes, will dominate the spectrum below 1 $\mu$m.

We now describe our simulations of JWST spectra of the planetary system
forward models described in \S\ref{sec:Systems} (Tables \ref{tbl-1} and
\ref{tbl-2}) and \S\ref{sec:Models}. As described in \citet{BBK14}, all
4 JWST science instruments have spectroscopic modes that will likely be
useful for observing transiting planets. Additional information on
expected instrument performance and spectroscopic observations of
transiting planets have been published for all instruments: NIRISS
\citep{DHB12}, NIRSpec \citep{FBB14, BKL15}, NIRCam \citep{GBE07}, and
MIRI \citep{KSB15}. The flexible capabilities of its instruments will
allow JWST users to often chose between more than one instrument mode
for obtaining a spectrum of a given wavelength range. As noted in
\S\ref{sec:Intro}, large spectral range, low-to-moderate spectral
resolution ($R \equiv \lambda / \delta \lambda \gtrsim 35$)
observations with high spectro-photometric precision have provided the
best scientific constraints on transiting exoplanet atmospheres to date.
Slitless spectra with good spatial sampling and stable detectors have
provided the best spectro-photometric precision \citep[e.g., HST WFC3
G141 mode as in][]{KBD14a}.

We have chosen to simulate $\lambda = 1 - 11$ $\mu$m transmission and
emission spectra of the modeled systems for the instrument modes listed
in Table~\ref{tbl-3}. We selected these modes because they best meet the
criteria noted above: large simultaneous wavelength coverage, adequate
spectral resolution ($R \gtrsim 100$), slitless operation, and bright
limits sufficient to observe the selected systems with high throughput.
They also have the best spatial sampling available on JWST over their
wavelength ranges; good sampling should minimize systematic errors due
to intrapixel response variations in the presence of pointing jitter
\citep[e.g.,][]{DSW09}. 
%We have ignored the small wavelength overlap
%between the NIRISS SOSS and NIRCam LW grism modes in order to simplify
%the assessment of what modes are important for constraining what
%scientific parameters. 
The NIRSpec instrument can also obtain spectra of
bright objects over the same wavelengths as NIRISS SOSS and the NIRCam LW grisms (also in 3 exposures), and its use in exoplanet transit
observations have been well studied \citep[e.g.,][]{FBB14, BKL15,
BBK14}. However, we chose to simulate NIRISS and NIRCam over $1 - 5$
$\mu$m wavelengths because of their slitless operation, finer spatial
sampling (64 vs 100 mas / pixel), and brighter flux limits (important
for the best targets).

\begin{deluxetable*}{lllrrl}
\tabletypesize{\scriptsize}
%\rotate
\tablecaption{Selected JWST Instrument Modes\label{tbl-3}}
\tablewidth{0pt}
\tablehead{
\colhead{Instrument} & \colhead{Mode} &  \colhead{Optics} &
\colhead{$\lambda (\mu$m)} & 
%\colhead{Native $R \equiv \lambda / (\delta \lambda)$} & \colhead{Comment}
\colhead{Native $R$\tablenotemark{a}} & \colhead{Sampling (pixels)\tablenotemark{b}}
}
\startdata
NIRISS		& bright SOSS & GR700XD &  1 -- 2.5\tablenotemark{c}	& $\sim  700$ & $\sim $25 \\
NIRCam		& LW grism & F322W2  &  2.5 -- 3.9	& $\sim 1700$ & $\sim$ 2 \\
NIRCam		& LW grism & F444W   &  3.9 -- 5.0	& $\sim 1700$ & $\sim$ 2 \\
MIRI		& SLITLESS & LRS prism & 5.0 -- 11\tablenotemark{d}	& $\sim 100$  & $\sim$ 2 
\enddata
\tablenotetext{a}{Spectral resolution $R \equiv \lambda / (\delta \lambda)$.}
\tablenotetext{b}{Spatial extent of point source spectrum in pixels.}
\tablenotetext{c}{Total NIRISS SOSS mode wavelength coverage spans $\lambda = 0.6 - 2.8$ $\mu$m, but we adopt the 1.0 $\mu$m short wavelength cutoff that is required for these bright stars and a 2.5 $\mu$m long wavelength cutoff to avoid spectral contamination.}
\tablenotetext{d}{We adopt a long-wavelength cutoff of 11 $\mu$m for the MIRI LRS because its transmission and SNR are degraded at longer wavelengths \citep[see Figs. 8 and 9 of][]{KSB15}.}
\end{deluxetable*}

\subsection{Signals and Random Noise Components}

Signals and noise of transmission and eclipse spectra were estimated for
each instrument mode (Table \ref{tbl-3}) for each planetary system
(Table \ref{tbl-1}). We combined these synthetic spectra to mimic the
observational sequence of observing a single transit or eclipse event
with each instrument mode (a total of 4 events for each transit or
eclipse), producing a complete simulated $\lambda = 1 - 11$ $\mu$m
spectrum and error bars estimating the uncertainty in each spectral
channel.

We computed signals for the separate stages of a transit or eclipse
event: star only or star + planet. Model stellar flux spectra were
constructed by interpolating Nextgen \citep{HAB99} models with surface
gravities and effective temperatures spanning the observed values for
each system. The in-transit star + planet flux was defined as $F(* +
p)_{Tr, \lambda} = F_{*, \lambda} \times (1 - (R_{p,\lambda}/R_*)^2)$.
The star + planet emission flux is simply the sum of the model star
flux and the model planet flux (see \S\ref{sec:Models}) at each
wavelength, $F(* + p)_{Em, \lambda} = F_{*, \lambda} + F_{p, \lambda}$.
Estimated detected signals of each of these astrophysical events were
computed for each system using the chosen system parameters (Table
\ref{tbl-3}), with

\begin{equation}
S_\lambda = F_\lambda A_{tel} t \frac{\lambda^2} {hcR} \tau
\end{equation}
where $S_\lambda$ is the signal per spectral element in electrons,
$F_\lambda$ is the star or star + planet spectral flux density, $t$ is
the total exposure time during the event, $\lambda$ is the wavelength of
the spectral element, $R$ is the resolution of the spectral bin,
$A_{tel}$ is the area of the JWST aperture (25 m$^2$), and $\tau$ is the
total system transmission (photon conversion efficiency) at that
wavelength. Total exposure time $t$ was calculated as the sum of the
photon collecting time of all integrations that are executed during 0.9
$\times$ the event duration ($T_{14}$ in Table \ref{tbl-2}). Individual
integration times were computed from the host star brightnesses (Table
\ref{tbl-2}) and the correlated double sample minimum integration times
and bright limits of each instrument mode \citep[c.f.][]{BBK14}, including the resulting readout efficiency.

We assumed that the star would also be observed for time $t$ before
and / or after the transit or eclipse event, yielding total
integration time $2t$ for the visit. The system transmission $\tau$
was computed to be the product of the 3 element JWST telescope (set to
0.9) and the selected instrument. NIRISS single object slitless
spectroscopy (SOSS) transmission $\tau$ was estimated to be 0.35 at
$\lambda = 1.25 \mu$m, falling off gradually at longer and shorter
wavelengths in an approximation of the first order blaze function of
its new GR700XD grism. Similarly, the NIRCam was estimated to have
$\tau = 0.30$ at the $3.7 \mu$m blaze peak of its grisms, falling off
at longer and shorter wavelengths in an approximation of the first
order grism blaze function \citep{GBE07}. The slitless photon
conversion efficiency curve ($\tau \simeq 0.3$) in Figure 9 of
\citep{KSB15} was adopted for the MIRI LRS $\tau$ values. \\

The JWST observatory is expected to have minimal natural (zodiacal) and
telescope background emission at near-IR wavelengths, but the telescope
background will increase quickly at wavelengths $\lambda \gtrsim 10
\mu$m. Background emission can become an important noise term when
observing faint stars with the MIRI LRS because its slitless mode
detects the full background of its $\lambda = 5 - 12 \mu$m bandpass. We
computed the background signal for each observation with the equation

\begin{equation}
Bkg = B t A_{pix} n_{pix} R_{native} / R
\end{equation}
where $Bkg$ is the background signal detected in each spectral bin, $B$
is the background of the instrument mode in electrons arcsec$^{-2}$ per
second, t is the total exposure time during the event, $A_{pix}$ is the
area subtended by each pixel in arcsec$^{2}$, $n_{pix}$ is the number of
spatial x 2 spectral pixels summed in each $R_{native}$ native 2-pixel
resolution element of the selected observing mode, and $R$ is the final
binned spectral resolution of the observation. The $B$ background values
were computed from the average zodiacal and telescope backgrounds of
each instrument mode provided by the STScI JWST prototype exposure time
calculator\footnote{http://jwstetc.stsci.edu/etc/}.

Noise values were computed at each wavelength as the sums of
photo-electron Poisson noise, background Poisson noise, and total (read
and dark) detector noise $N_{d,tot}$:

\begin{equation}
N_\lambda = \sqrt{S_\lambda + Bkg + N_{d,tot}^2}
\end{equation}

where 

\begin{equation}
%N_{d,tot} = N_{d}  \sqrt(n_{pix})  \sqrt(n_{ints}) \sqrt(R_{native} / R )
N_{d,tot} = N_{d}  \sqrt{n_{pix} n_{ints} R_{native} / R}.
\end{equation}

$N_{d}$ is the total (read and dark current) detector noise of a
single integration, and $n_{ints}$ is the number of integrations during
exposure time $t$. We set $N_{d}$ = 18 electrons for the HgCdTe
detectors (NIRISS and NIRCam modes) and $N_{d}$ = 28 electrons for MIRI;
these correlated double sample noise values are expected maxima for
nearly all observations. The numbers of integrations $n_{ints}$ were set
using the bright limits, efficiencies, and frame times of each
instrument mode given in \citet{BBK14} such that the total real time fit
within 0.9 times the transit duration (T$_{14}$ in Table~\ref{tbl-2}),
summing to total exposure time $t$ for the event (plus equal time on the
host star alone).\\

\subsection{Final simulated spectra} \label{sec:final_spectra}

The signals and backgrounds of the astrophysical events were combined as follows to estimate the final transmission ($Tr_\lambda$) and emission ($Em_\lambda$) spectra:
\begin{equation}
Tr_\lambda = \frac{S_\lambda(F_{*}) + Bkg - (S_\lambda(F(* + p)_{Tr}) + Bkg)} 
{S_\lambda(F_{*}) + Bkg - \langle Bkg \rangle}
\end{equation}

and

\begin{equation}
Em_\lambda = \frac{S_\lambda(F(* + p)_{Em}) + Bkg - (S_\lambda(F_{*}) + Bkg)} 
{S_\lambda(F_{*}) + Bkg - \langle Bkg \rangle}.
\end{equation}

Note that $Tr_\lambda \equiv (R_{p,\lambda}/R_{*})^2$ and $Em_\lambda
\equiv F_{p, \lambda} / F_{*}$ in noiseless cases. These $Tr_\lambda$
and $Em_\lambda$ signals were computed for the chosen system,
instrument, and observation parameters for each planetary model in
Table~\ref{tbl-1}. We binned spectral channels to achieve a final
resolution of $R \leq 100$ per wavelength bin; instrument modes with
higher intrinsic $R$ were binned to $R=100$, while ones with lower $R$
\citep[i.e., MIRI LRS at $\lambda < 8 \mu$m; see ][]{KSB15} were not
binned. We chose $R=100$ as a compromise value low enough to maximize
signal-to-noise while also being high enough to resolve molecular band
spectral features. 

We did perform a preliminary test of how binning impacts the
retrieved uncertainties of H$_2$O, CO, and CO$_2$ in the hot Jupiter
solar composition cloudy transmission spectrum. We found that binning
$\lambda = 1 - 5 \mu$m (NIRISS + NIRCam) spectra to $R=350$ (2 NIRISS
resolution elements) did not produce any mixing ratio uncertainties that
were less than ones retrieved for data binned to $R=100$ in 2 trials
using different instances of $N_{d,tot}$ (no systematic noise was
included). However, the ideal binning of each observing mode will
likely be sensitive to actual in-flight noise performance, and this will
not be known until after JWST begins operations.

The large aperture of JWST will ensure that the observatory will
collect a large number of photons on bright stars, potentially
resulting in the detection of $\sim 10^{10}$ or more photo-electrons
per spectral bin when summing a significant number of observations of
transit or eclipse events. In such cases, stellar photon noise will
dominate $N_\lambda$, resulting in signal-to-noise ratios SNR $\sim
10^5$ or $N_\lambda$ values being only ~10 parts per million (10 ppm)
of signal values. Existing observations of transiting planets have not
yielded such low noise values. Either astrophysical noise
\citep[e.g.,][]{BAI15} and / or instrumental noise (e.g.,
de-correlation residuals) produce systematic noise floors that are not
lowered when summing more data. The best HST WFC3 G141 observations of
transiting systems to date have noise on the order of 30 ppm
\citep{KBD14a}, while observations of transiting planets with the
Spitzer Space Telescope Si:As detectors showed noise as low as $\sim
65$ ppm \citep{KCC09}.

We adopt reasonably optimistic systematic noise floor values of 20
ppm, 30 ppm, and 50 ppm for NIRISS SOSS ($\lambda = 1 - 2.5 \mu$m),
NIRCam grism ($\lambda = 2.5 - 5.0 \mu$m), and MIRI LRS ($\lambda =
5.0 - 11 \mu$m) observations, respectively. These are less than or
equal to the values estimated by \citet{DSW09} for the JWST NIRSpec
and MIRI instruments. The excellent spatial sampling of the NIRISS
GR700XD SOSS grism approaches that of the HST WFC3 G141 spatial
scanning mode, and both instruments have reasonably similar HgCdTe
detectors. We anticipate that de-correlation techniques will continue
to improve, so we assign a 20 ppm noise floor value to NIRISS even
though HST has not yet done quite this well. NIRCam will have similar
detectors but will not be sampled as well (see Table~\ref{tbl-3}), so
we assign it a systematic noise floor equivalent to HST's best
performance to date. The MIRI imager / LRS detector is similar (in
materials, architecture, and sampling) to the Spitzer IRAC Si:As
detector used by \citet{KCC09}, and we use this as the basis for
assigning a 50 ppm noise floor to MIRI. We will not know the actual
performance of these instruments until after JWST commissioning, but
\citet{BBK14} have demonstrated decorrelated noise precision similar
to the adopted NIRISS and NIRCam values in laboratory tests devised to
simulate their observations of transiting planets.

\begin{figure*}
\begin{center}
\includegraphics[width=1.0\textwidth, angle=0]{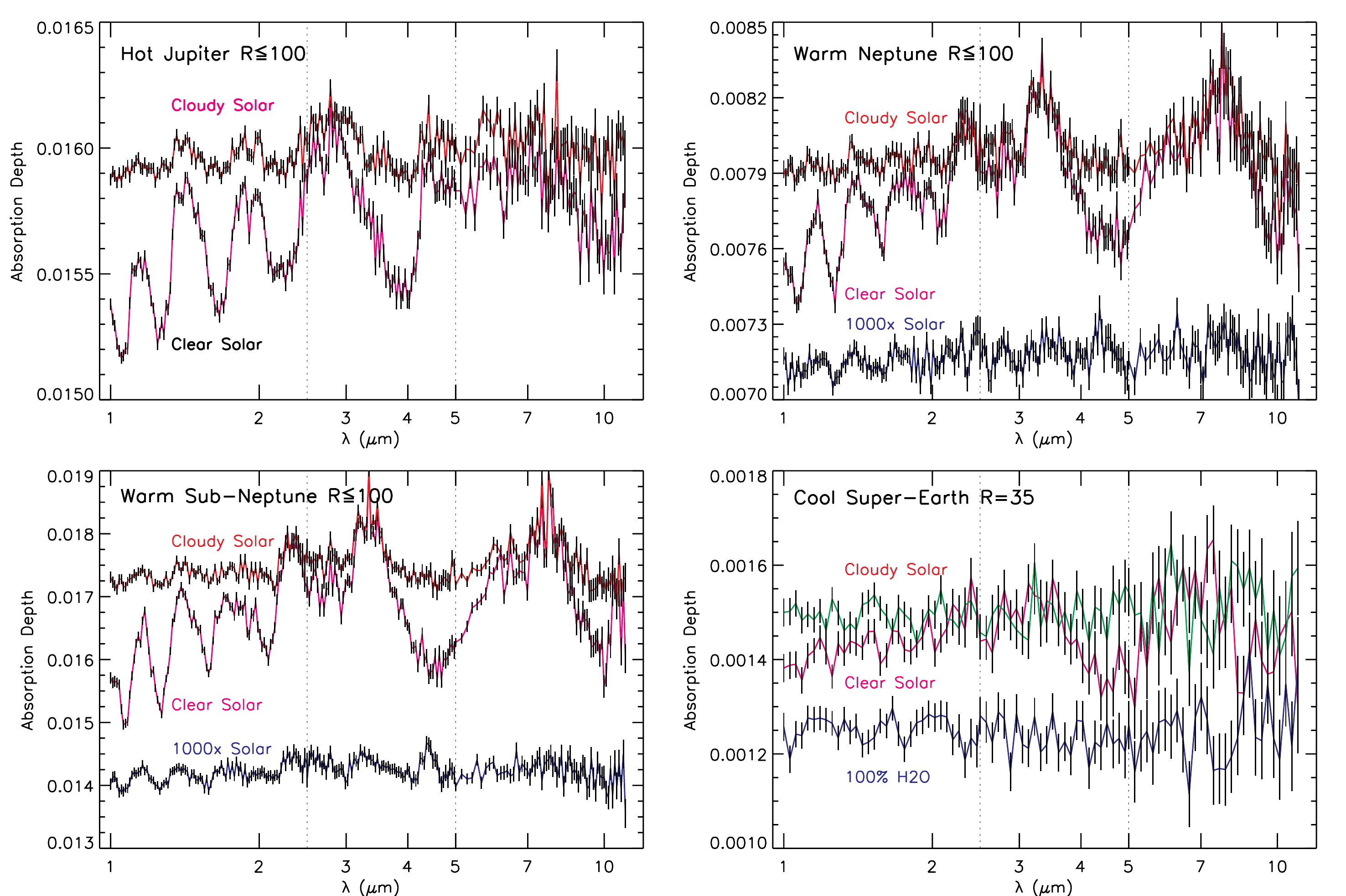}
\end{center}
     \caption{ \label{fig:WNTr_spec_fig} 
Final simulated transmission spectra of the transmission models shown in Fig. \ref{fig:forward_trans_spec_fig}.
The $Tr_\lambda = (R_{p,\lambda}/R_{*})^2$ spectra 
are for a single transit with equal time on the star alone for each of the 4 instrument modes (Table~\ref{tbl-3}). Spectra have been been binned to resolution $R \leq 100$  (hot Jupiter, warm Neptune, warm sub-Neptune; see \S\ref{sec:final_spectra}) as used for all retrievals or $R = 35$ (cool super-Earth) for display purposes only.
The simulated spectra include a noise instance and are presented as colored curves. The black error bars denote $1 \sigma$ of noise composed of random and systematic components. Dashed lines show the wavelength range boundaries of the chosen NIRISS, NIRCam, and MIRI instrument modes.
  }
\end{figure*}

\begin{figure*}
\begin{center}
\includegraphics[width=1.0\textwidth, angle=0]{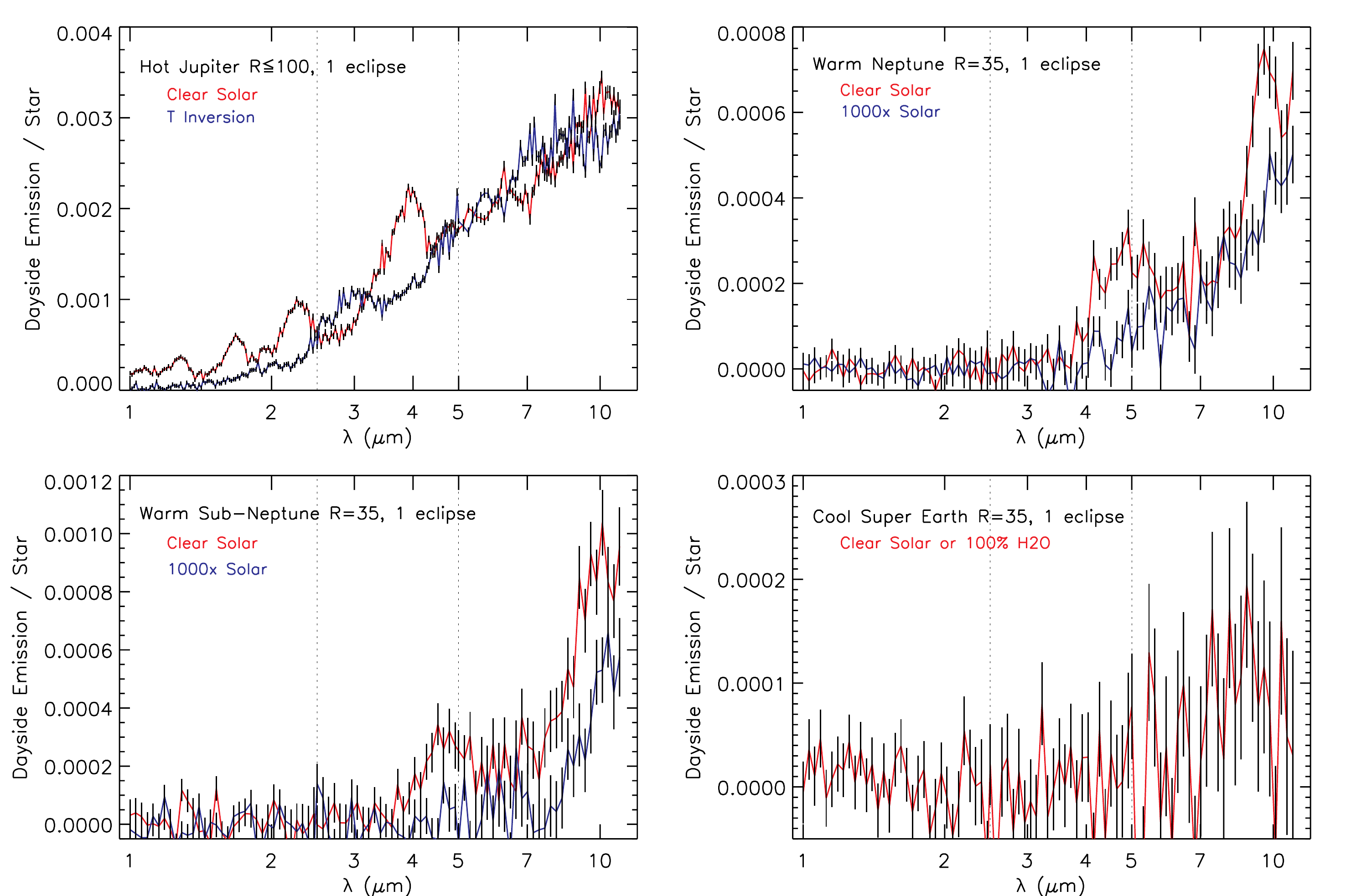}
\end{center}
     \caption{ \label{fig:WNTr_emis_spec_fig} 
Final simulated spectra of the emission models shown in Fig.
\ref{fig:forward_emis_spec_fig}. The $Em_\lambda = S(F_p)/S(F_{*})$
spectra are for a single eclipse with equal time on the star alone for each of the 4 instrument modes (Table~\ref{tbl-3}).
Spectra have been been binned to resolution $R \leq 100$ as used for all retrievals (hot Jupiter; see \S\ref{sec:final_spectra}) or $R = 35$ for display purposes only (warm Neptune, warm sub-Neptune, cool super-Earth).  The simulated spectra include a noise instance and are presented as colored curves. The black error bars denote $1 \sigma$ of noise composed of random and systematic components. Dashed lines show the wavelength range boundaries of the chosen NIRISS, NIRCam, and MIRI instrument modes. Only 1 model is shown for the cool super-Earth for clarity; the noise is much greater than the difference between the clear solar and 100\% H$_2$O atmospheres.
  }
\end{figure*}

A single instance of noise was computed for each transmission or
emission observation, using the expression for $N_\lambda$ and
propagating through the relation $Tr_\lambda$ or $Em_\lambda$. We then
added the appropriate systematic noise floor in quadrature, using the
values given above. The single noise instance was drawn from a
distribution with this amplitude, and the instance was added to the
$Tr_\lambda$ or $Em_\lambda$ signal to produce the final simulated
spectrum for each planet model in Table~\ref{tbl-1}. Uncertainties
were set to the total noise amplitude determined for each bin; Figures
\ref{fig:WNTr_spec_fig} and \ref{fig:WNTr_emis_spec_fig} show the
simulated spectra for each case with these uncertainties plotted as
error bars for 1 example. We computed different noise instances for
each instrument mode (NIRISS SOSS, NIRCam LW grism, and MIRI slitless
LRS). The resultant spectra and uncertainties were then used to
retrieve the parameters of each model planet's atmosphere as discussed
in \S\ref{sec:Models}.

%%%%%%%%%%%%% figure %%%%%%%%%%%%%%%%%%%%
%%%%%%%%%%%% figure %%%%%%%%%%%%%%%%%%%%

%%%%%%%%%%%%% figure %%%%%%%%%%%%%%%%%%%%
%%%%%%%%%%%% figure %%%%%%%%%%%%%%%%%%%%

\section{Retrieval Results} \label{sec:Results}

The retrievals were performed for the wavelength ranges of up to 3 different
instrument combinations for each of the model planet atmospheres: NIRISS
($\lambda = 1.0 - 2.5$ $\mu$m), NIRISS + NIRCam ($\lambda = 1.0 - 5.0$
$\mu$m), and NIRISS + NIRCam + MIRI LRS ($\lambda = 1.0 - 11$ $\mu$m).
Retrievals were performed for all 3 combinations in cases of high
signal-to-noise, i.e. the transmission spectra of all planet atmospheres and
the hot Jupiter emission spectra. The warm Neptune and warm sub-Neptune
emission spectra had insufficient signal-to-noise for retrievals with only
NIRISS ($1.0 - 2.5$ $\mu$m) data, and {\it the complete $\lambda = 1 - 11$
$\mu$m cool super-Earth emission spectrum had insufficient signal-to-noise for
retrievals}. There simply is not enough flux contrast $F_p/F_*$ for
useful emission spectra from this system when a single secondary eclipse is
observed at each wavelength. Photometric filter observations may be more
useful for constraining the planet's properties. Small ($R\lesssim 2
R_\oplus$), cool ($T<$700 K) planets will need host stars with $K \lesssim
8.5$ mag and / or spectral types later than M0 V for useful emission spectra
of single secondary eclipses.

In this section, we focus on the retrieved parameters that most
directly impact the simulated spectra: mixing ratios of significant
molecular absorbers (CH$_4$, CO, CO$_2$, H$_2$O, NH$_3$), clouds, and
atmospheric temperature-pressure (T-P) profiles. C/O (carbon-to-oxygen
ratios) and [Fe/H] are {\em derived} (not retrieved) from these
quantities by a Monte-Carlo propagation of the molecular uncertainties
as in \citet{LWZ13} and \citet{LTB15}, respectively.

\begin{figure*}
\begin{center}
\includegraphics[width=1.0\textwidth, angle=0]{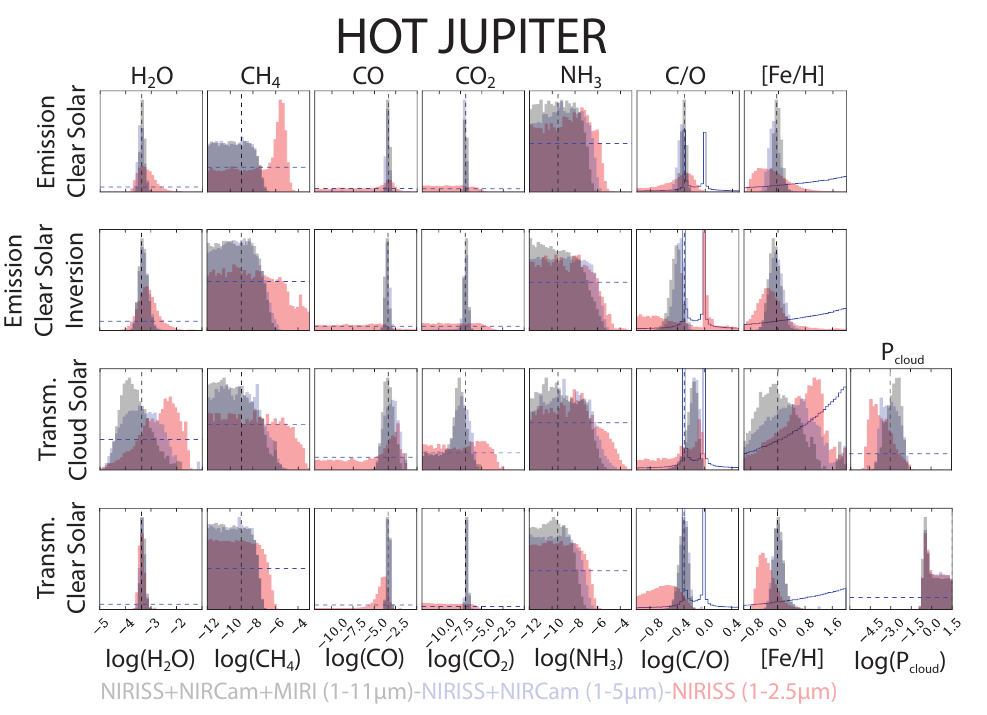}
\end{center}
     \caption{ \label{fig:HJ_gasses_fig} 
Retrieved and derived quantities of the hot Jupiter planet. Retrieved volume mixing ratios are shown for each molecular species. Marginalized posterior histogram shadings are color coded by instrument mode and therefore wavelengths of spectra used for retrievals. Priors are indicated by blue horizontal dashed lines for the retrieved molecules. The resulting distributions for the derived quantities log (C/O) and [Fe/H] from the gas priors are more complex distribution functions (shown in blue). True values are indicated by vertical dashed black lines for all quantities. 
  }
\end{figure*}

\begin{figure*}
\begin{center}
\includegraphics[width=1.0\textwidth, angle=0]{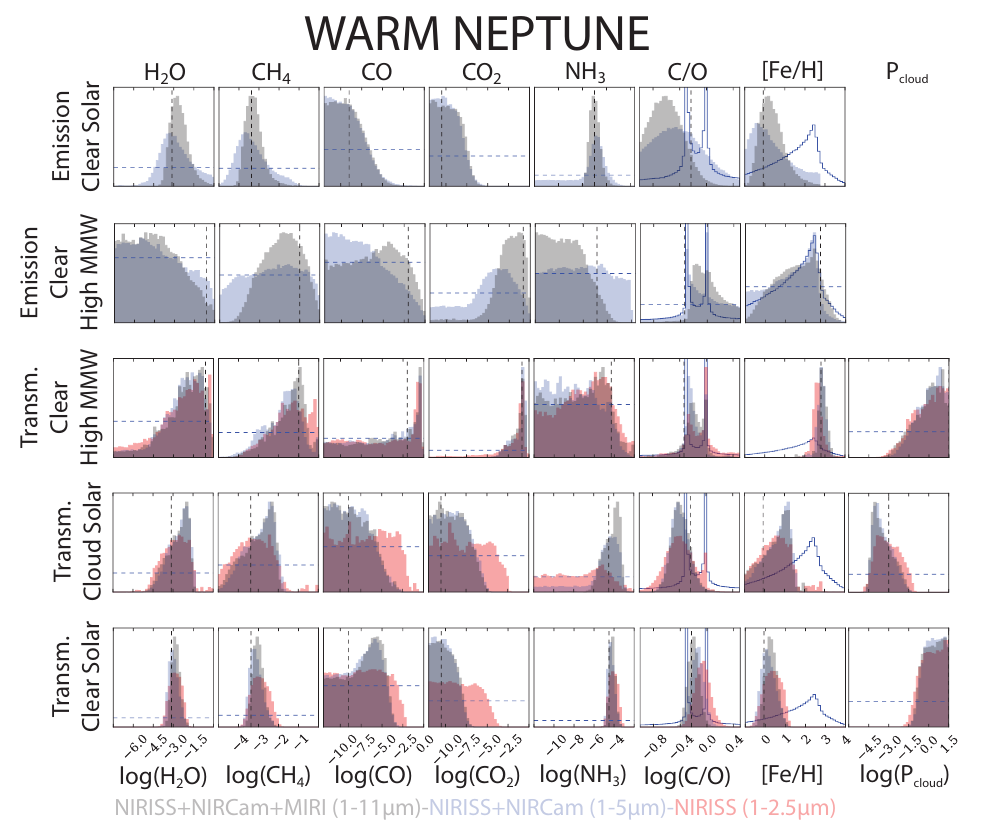}
\end{center}
     \caption{ \label{fig:WN_gasses_fig} 
Retrieved and derived quantities of the warm Neptune planet. Retrieved mixing ratios are shown for each molecular species. Marginalized posterior histogram shadings are color coded by instrument mode and therefore wavelengths of spectra used for retrievals. Priors are indicated by blue horizontal dashed lines for the retrieved molecules. The resulting distributions for the derived quantities log (C/O) and [Fe/H] from the gas priors are more complex distribution functions (shown in blue). True values are indicated by vertical dashed black lines for all quantities. 
   }
\end{figure*}

\begin{figure*}
\begin{center}
\includegraphics[width=1.0\textwidth, angle=0]{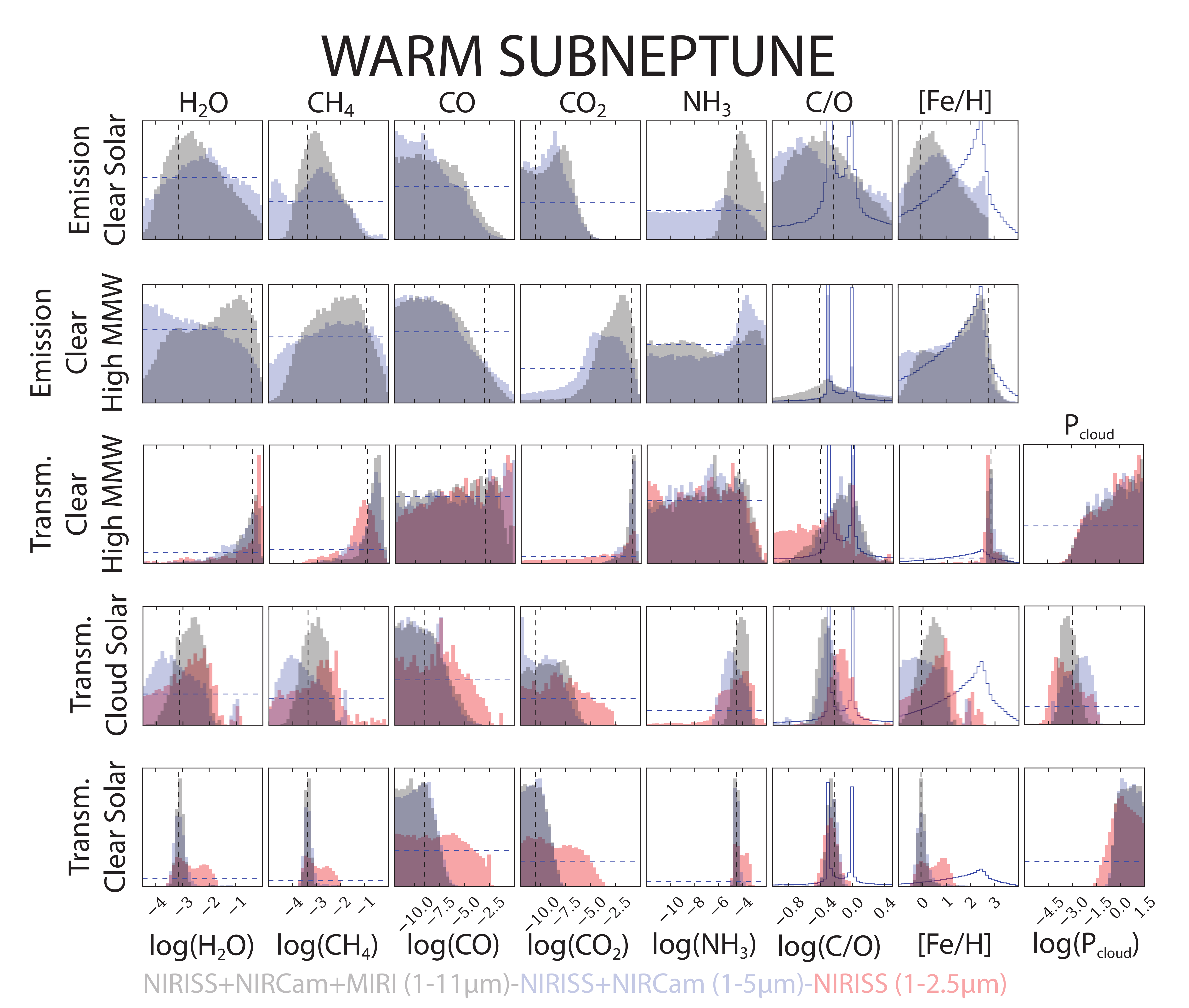}
\end{center}
     \caption{ \label{fig:WSN_gasses_fig} 
Retrieved and derived quantities of the warm sub-Neptune planet. Retrieved mixing ratios are shown for each molecular species. Posterior predictive histogram shadings are color coded by instrument mode and therefore wavelengths of spectra used for retrievals. Priors are indicated by blue horizontal dashed lines for the retrieved molecules. The resulting distributions for the derived quantities log (C/O) and [Fe/H] from the gas priors are more complex distribution functions (shown in blue). True values are indicated by vertical dashed black lines for all quantities. 
  }
\end{figure*}

\begin{figure*}
\begin{center}
\includegraphics[width=1.0\textwidth, angle=0]{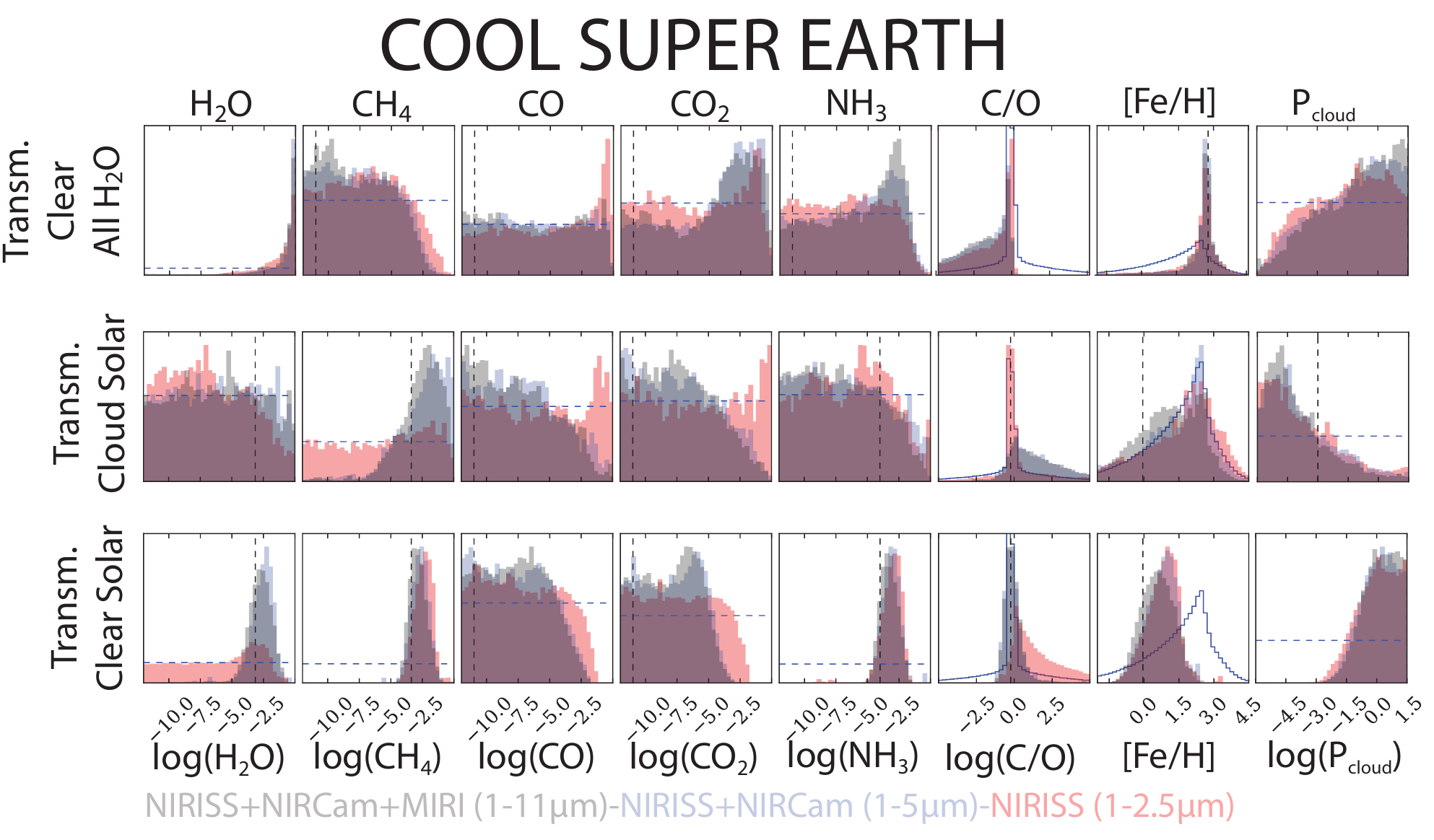}
\end{center}
     \caption{ \label{fig:CSE_gasses_fig} 
Retrieved and derived quantities of the cool super-Earth planet. Retrieved mixing ratios are shown for each molecular species. Posterior predictive histogram shadings are color coded by instrument mode and therefore wavelengths of spectra used for retrievals. Priors are indicated by blue horizontal dashed lines for the retrieved molecules. The resulting distributions for the derived quantities log (C/O) and [Fe/H] from the gas priors are more complex distribution functions (shown in blue). True values are indicated by vertical dashed black lines for all quantities. 
  }
\end{figure*}

Figures \ref{fig:HJ_gasses_fig} -- \ref{fig:CSE_gasses_fig} summarize
the marginalized posteriors for the relevant retrieved parameters and
the derived C/O and [Fe/H] for the different planet and atmosphere
scenarios. The input forward model true values and retrieval priors are
also indicated in the figures. Any offsets between retrieved
distribution medians and true values are due to the particular instance
of random noise on the simulated spectra. Note that the input [Fe/H]
values are slightly lower than the solar ([Fe/H]=0) and
1000$\times$ solar ([Fe/H]=3) bulk values that were used to construct
the forward models (see \S\ref{sec:Systems}). This is because some of
the O atoms have been removed to account for the expected formation of
Mg$_2$SiO$_4$ (enstatite) condensates in clouds in deep atmospheres
below the regions probed by emission or transmission spectra (except in
the Hot Jupiter)\footnote{Also, the true metallicities for the emission
spectra are slightly lower than for transmission as we did not include
N$_2$ in the total metallicity calculation}.

% Also, [Fe/H] values were computed from mixing ratios at a chosen temperature
% and not integrated over the whole profile, but no need to say that here 
% unless asked 

%%%%%%%%%%%%% figure %%%%%%%%%%%%%%%%%%%%
%%%%%%%%%%%% figure %%%%%%%%%%%%%%%%%%%%

%%%%%%%%%%%%% figure %%%%%%%%%%%%%%%%%%%%
%%%%%%%%%%%% figure %%%%%%%%%%%%%%%%%%%%

%%%%%%%%%%%%% figure %%%%%%%%%%%%%%%%%%%%
%%%%%%%%%%%% figure %%%%%%%%%%%%%%%%%%%%
%\newpage

%%%%%%%%%%%%% figure %%%%%%%%%%%%%%%%%%%%
%%%%%%%%%%%% figure %%%%%%%%%%%%%%%%%%%%

The prior distributions on the retrieved parameters are
uniform-in-log. However, uniform-in-log priors bias the resulting C/O
and [Fe/H] distributions \citep[see discussion in][]{LWZ13}, thus we
show what these distributions look like resulting from the
uniform-in-log priors on the gas mixing ratios. These would be the
distributions obtained had we made no observations. Therefore these
prior-derived distributions must be considered in interpreting the
posterior log (C/O) and [Fe/H] distributions.

\subsection{Impacts of Atmospheric Parameterization}\label{sec:Parameterization} 

We now investigate whether the chosen atmospheric parameterization and priors may impact the retrieved results. \citet{KLB15} used both the ``free" and ``chemically consistent" retrieval approaches to determine the C/O in the transmission spectrum of WASP-12b. They found consistent results amongst the two approaches suggesting a robust solution. We perform a similar analysis here. In addition to retrieving the mixing ratios of the individual molecules independent of the chemistry that links them (the ``free" approach performed thus far), we also retrieve log C/O and [Fe/H] directly as a test case (with prior ranges from -2 to 2, and -4 to 4 respectively). These quantities, together with the temperature and pressure at each level in the atmosphere, permit the determination of the thermochemical equilibrium gas mixing ratios.  These solutions are advantageous as they are chemically plausible and thermochemically self-consistent.  However, such an approach does not account for the nearly limitless possible combinations of physical and chemical processes occurring in planetary atmospheres (e.g., vertical mixing, photochemistry, ion chemistry, 3-D transport, cloud-gas microphysics interactions, interior-atmosphere chemistry coupling, escape processes, non-LTE, etc.) so we have not adopted it as our primary technique.  

\begin{figure}
\begin{center}
\includegraphics[width=0.45\textwidth, angle=0]{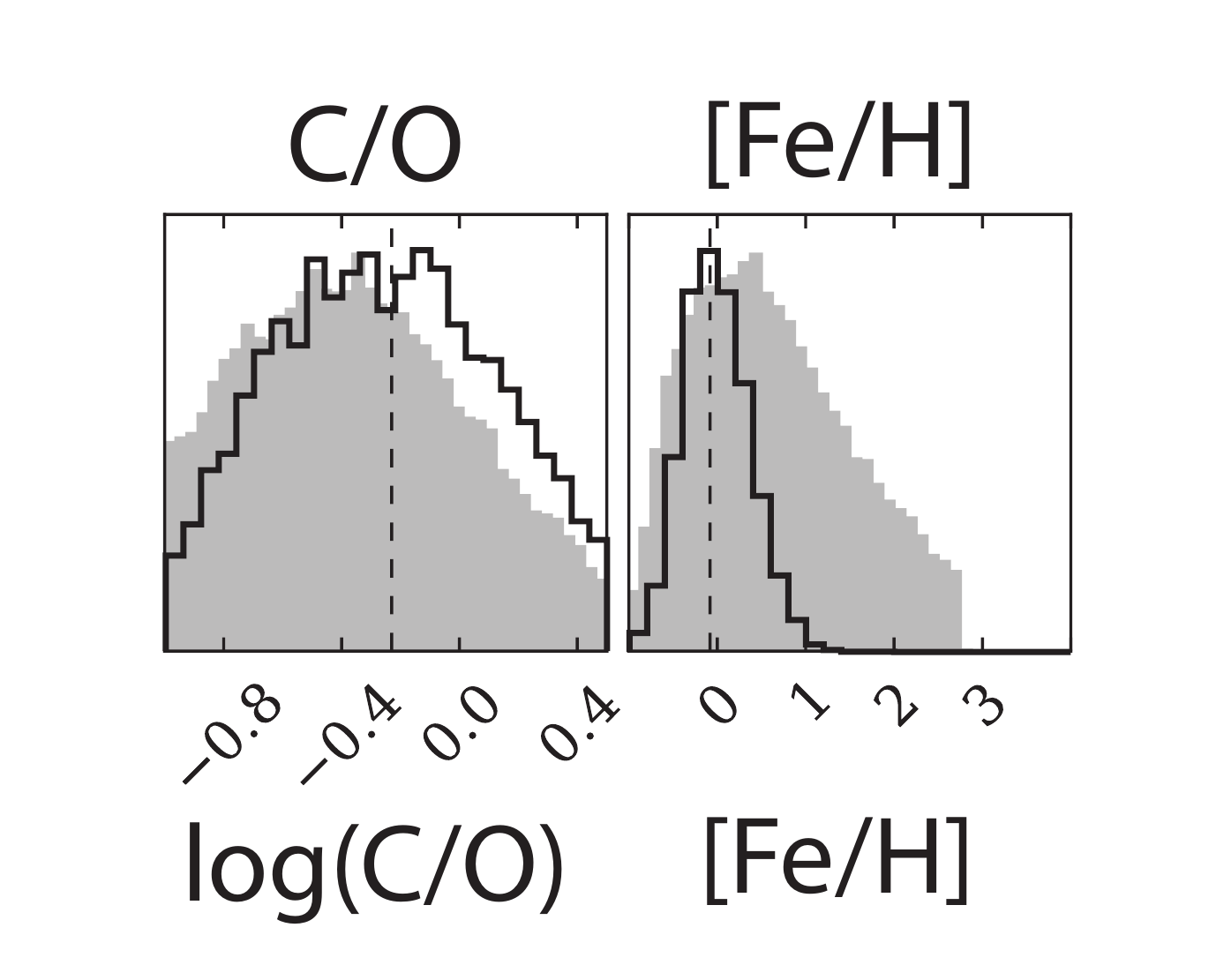}
\end{center}
     \caption{  \label{fig:dist_comparison}
Comparison between the classic ``free" retrieval approach used in this work and the ``chemically consistent" approach for the $\lambda = 1 - 11$ $\mu$m emission spectrum of the warm sub-Neptune solar composition atmosphere.  The gray solid histograms show the same values as in the top row of Fig.~\ref{fig:WSN_gasses_fig} ($1 - 11$ $\mu$m wavelengths).  The black unfilled histograms are from the thermochemically-consistent retrieval.  The log (C/O) histograms are in good agreement, both indicating poor constraint of C/O for this planetary atmosphere. However, the [Fe/H] metallicity histograms are substantially different due to the imposition of thermochemical consistency. }
\end{figure}

While not comprehensive, we perform one example using this chemically
consistent approach and examine the impact it has on our ability to
infer the atmospheric metallicity and C/O ratio. We retrieve on the
full $1 - 11$ $\mu$m wavelength warm sub-Neptune emission spectrum for
the clear solar composition atmosphere.
Figure~\ref{fig:dist_comparison} illustrates the resulting marginalized
log(C/O) and [Fe/H] metallicity posteriors compared with those {\it
derived} from the molecular abundances in our baseline free retrievals
(i.e., Fig.~\ref{fig:WSN_gasses_fig}, top row). The C/O histograms are
qualitatively similar; both suggest a weak constraint of the C-to-O
ratio. Perhaps more interesting is the comparison of the metallicity
histograms. The chemically consistent approach provides a
several-orders-of-magnitude better metallically constraint than that
derived from retrieving the molecular mixing ratios freely. This is
because the chemically consistent approach rules out combinations of
molecular abundances that do not abide by thermochemical equilibrium.
Effectively, more prior information is being added to the chemically
consistent retrieval system in the form of a more sophisticated
parameterization with more assumptions but with fewer free parameters.
More generally, it would be possible to apply chemically consistent
models on all posterior ``free'' retrieval histograms (Figures \ref{fig:HJ_gasses_fig} -- \ref{fig:CSE_gasses_fig}) to rule out
non-physical parameter spaces within the equilibrium framework. This
would be an intermediate step between the classic retrieval and forced
self-consistency, but we do not implement it here.

Given a high enough signal-to-noise ratio and sufficient spectral resolution along with correct physics and chemistry constraints, one would expect the two approaches to produce the same distributions of the retrieved quantities. That would suggest true independence from any prior assumptions, and this would be the ideal regime for learning more about these atmospheres.

%%%%%%%%%%%%% figure %%%%%%%%%%%%%%%%%%%%
%%%%%%%%%%%% figure %%%%%%%%%%%%%%%%%%%%

\subsection{Temperature-Pressure Profiles and Parameter Uncertainties}

\begin{figure*}
\begin{center}
\includegraphics[width=0.83\textwidth, angle=0]{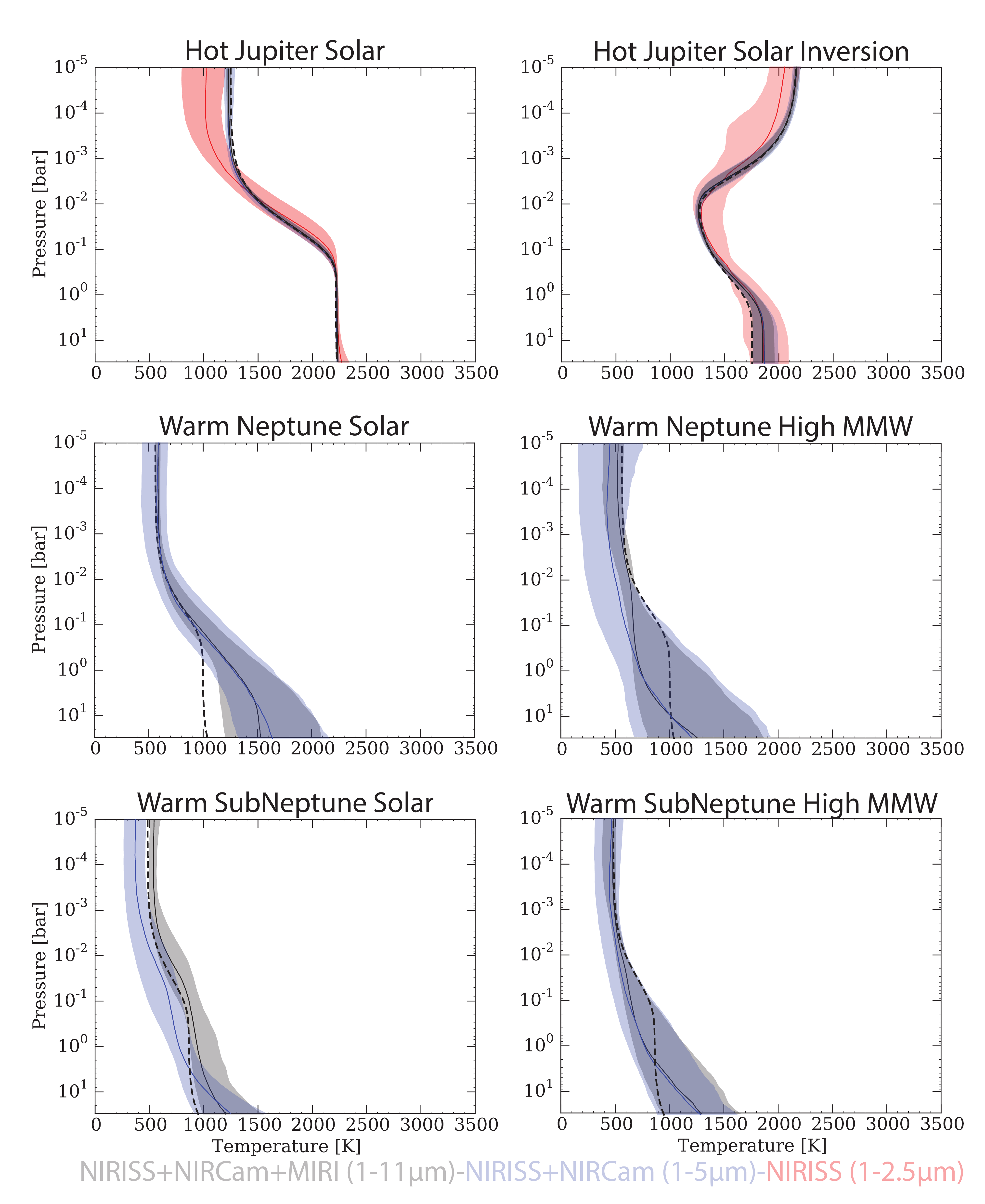} %Bigger blows up
\end{center}
     \caption{  \label{fig:TP_summary_fig}
Temperature-Pressure (T-P) profiles retrieved from emission spectra. True values are shown as dashed lines. Solid lines are best fit retrieved values, and shaded regions denote 1$\sigma$ uncertainties. Note that shadings are color coded by instrument mode and therefore the wavelengths of spectra used for retrievals. Sections of the
temperature pressure profiles outside of well-constrained regions (see text and Fig.~\ref{fig:TP_contrib_func}) are extrapolated from the temperature profile parameterization.
  }
\end{figure*}

\begin{figure*}
\begin{center}
\includegraphics[width=0.49\textwidth, angle=0]{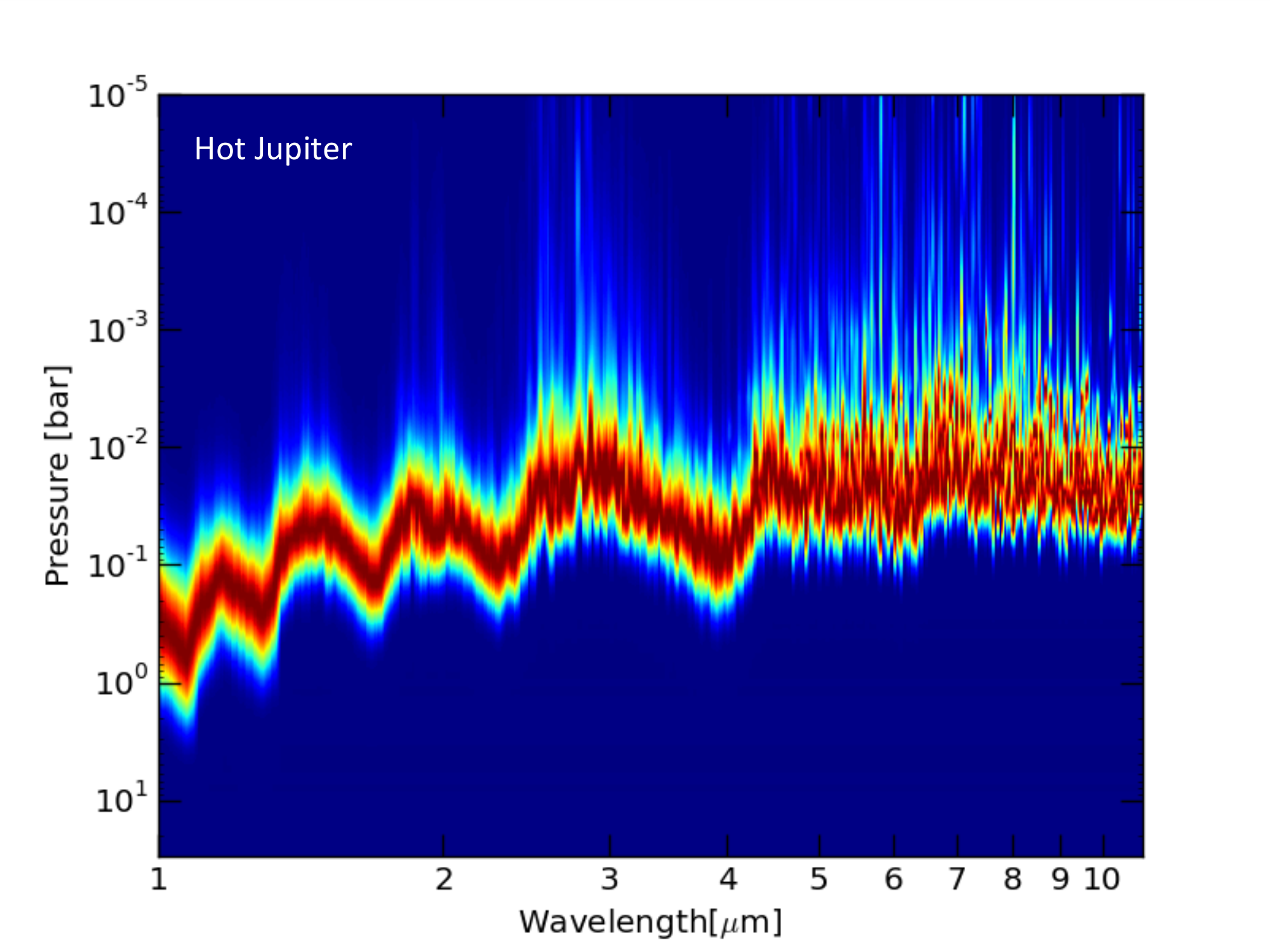} 
\includegraphics[width=0.49\textwidth,
angle=0]{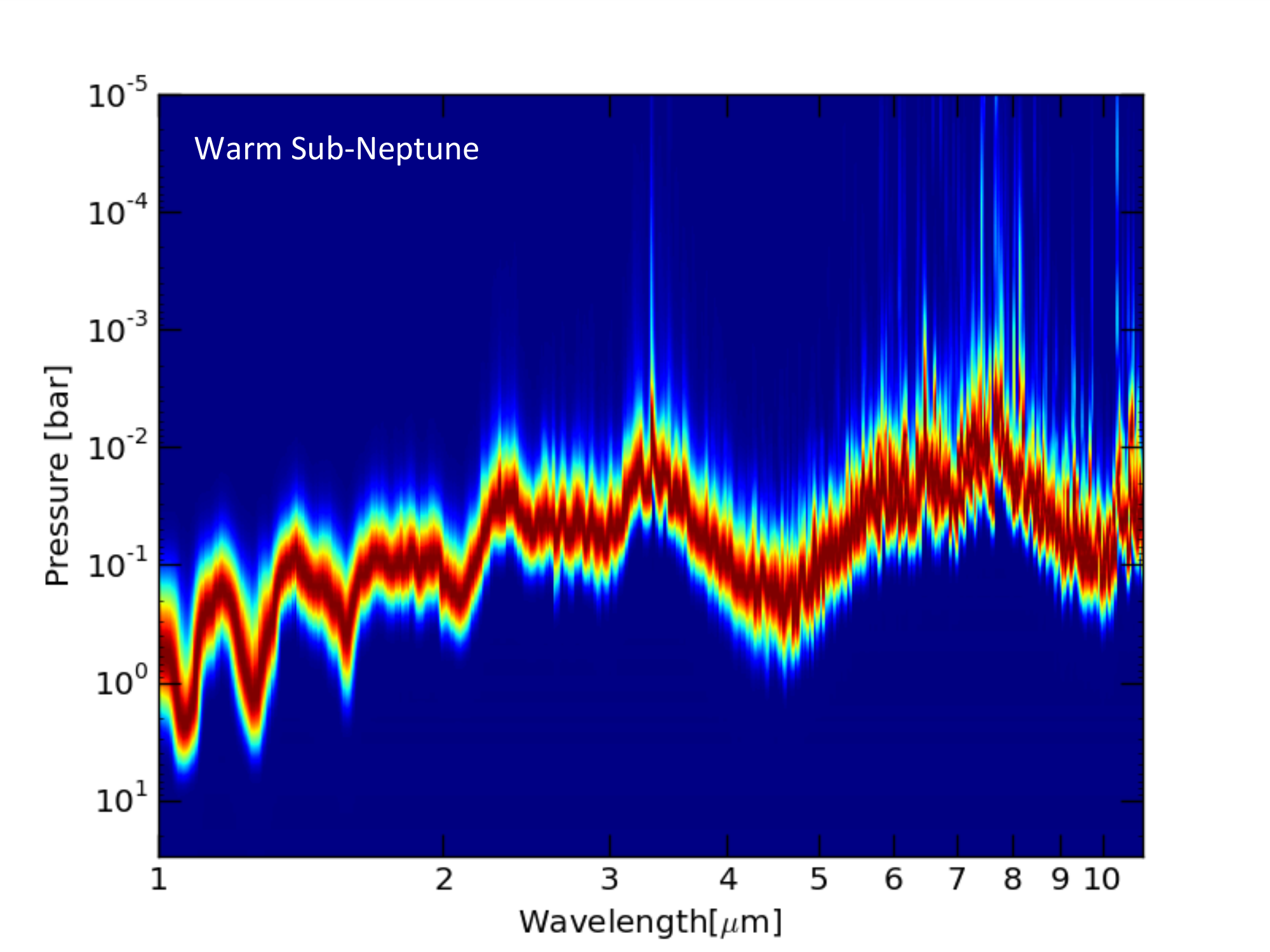} 
\end{center}
     \caption{   \label{fig:TP_contrib_func}
Normalized thermal emission contribution functions for the solar composition hot Jupiter (left) and warm sub-Neptune (right) planets. The figure shows the derivative of transmittance with altitude weighted by the planck function at that level. Dark red areas are where the emission predominantly originates whereas blue areas represent little or no emission. The pressure levels probed by the red areas contribute most to the retrieved composition and thermal structure information \citep[e.g., see][]{LWZ13,LKW14}.  The warm Neptune solar composition contribution function is very similar to the warm sub-Neptune one. }
\end{figure*}

Figure~\ref{fig:TP_summary_fig} shows the range of T-P profiles
retrieved from the simulated emission spectra over the 3 different
wavelength ranges. Figure~\ref{fig:TP_contrib_func} shows
normalized thermal emission contribution functions for the solar
composition hot Jupiter and warm sub-Neptune planets to illustrate
where their thermal emissions originate. The solar composition warm
Neptune contribution function is very similar to the warm sub-Neptune
one.

%%%%%%%%%%%%% figure %%%%%%%%%%%%%%%%%%%%
%%%%%%%%%%%% figure %%%%%%%%%%%%%%%%%%%%

%%%%%%%%%%%%% figure %%%%%%%%%%%%%%%%%%%%
%%%%%%%%%%%% figure %%%%%%%%%%%%%%%%%%%%

Table \ref{tbl-4} lists the the 68\% confidence intervals of the retrieved parameters of the transmission and emission scenarios (see \S\ref{sec:Models}, Table \ref{tbl-1}). For cases in which a molecule is not particularly abundant (e.g., CH$_4$ in the hot Jupiter and CO or CO$_2$ in the cooler objects), only upper limits could be obtained. We quote the 3$\sigma$ upper limits instead of confidence widths in these case. We caution however, that many of these upper limits are relatively soft and that one should really look at the histograms to get a sense for the distribution.

These numbers are meant to be a guide to illustrate how the constraints change from one object to the next. We also note that there is typically a $\sim$10$\%$ uncertainty on these uncertainty values due to the particular instances of random noise applied. In the remainder of this section, we examine how well the directly retrieved parameters constrain the different planetary atmospheres.

%\newpage

\subsection{Transmission Spectra Constraints}

Figures~\ref{fig:HJ_gasses_fig} -- \ref{fig:CSE_gasses_fig} and
Table~\ref{tbl-4} show that the mixing ratios of the dominant
molecules for a particular planet type are well constrained (bounded
posterior rather than an upper limit) in the transmission spectra of
all clear solar atmospheres in most cases. Furthermore, observations
with NIRISS alone ($\lambda \leq 2.5$ $\mu$m) can constrain H$_2$O in
these atmospheres nearly as well as observations out to 11 $\mu$m
wavelengths in some cases. For example, Table \ref{tbl-4} shows that
the retrieved H$_2$O mixing ratio of the clear solar hot Jupiter has a
68\% uncertainty of $0.16$ dex (45\%) for a $\lambda = 1 - 11$ $\mu$m
spectrum, and this degrades to only 0.20 dex (58\%) when only using
the $1 - 2.5$ $\mu$m wavelengths in a NIRISS observation. However, CO
is well constrained (uncertainty $\sim0.3$ dex for retrievals done
with the full $1 - 11$ $\mu$m or $1 - 5$ $\mu$m spectra, but the
uncertainty is much worse ($\sim1.8$ dex) when only the NIRISS data
are used.

The H$_2$O and CH$_4$ uncertainties of the warm Neptune clear solar
atmosphere increase by $\sim$0.3 dex (factor of $\sim$2), providing
uncertainties of $\sim$1 dex (factor of $\sim$10) when using only the
NIRISS wavelengths. These species have similar NIRISS-only
uncertainties in the warm sub-Neptune clear solar atmosphere, but
their full $\lambda = 1 - 11$ $\mu$m spectra provide better
uncertainties ($\sim$0.2 vs $\sim$0.5 dex for the warm Neptune). These
values are about $\sim$1.5 dex for the full $1 - 11$ $\mu$m
wavelengths for the cool super-Earth clear solar atmosphere, and the
CH$_4$ uncertainty does not change much when only considering the
NIRISS data range. However, the retrieved H$_2$O mixing ratio value
becomes an upper limit. Overall it is encouraging that these
atmospheres can be constrained so well in these modest observations (1
transit at each wavelength), and the wavelength range of NIRISS alone
is sufficient to detect dominant species with good confidence or even
measure mixing ratios to good precision in some cases (i.e., the
chosen hot Jupiter). Even for gases that are not abundant in the
atmospheres (CH$_4$ and NH$_3$ in the hot Jupiter, CO and CO$_2$ in
the cooler objects) relatively low upper limits can be obtained. For
instance, most scenarios within the hot Jupiter planet type can
detect, or rather, rule out NH$_3$ and CH$_4$ at the $\sim1$ ppm level.

Planets with cloudy atmospheres will be more challenging to constrain
with transmission spectra; this is already apparent from numerous HST
observations of GJ 1214b \citep[e.g.,][and references therein]{KBD14a,
CAJ11} and other planets that exhibit weak or no spectral features.
Figures \ref{fig:HJ_gasses_fig} -- \ref{fig:CSE_gasses_fig} and Table
\ref{tbl-4} show that the mixing ratios of most molecules are not
constrained well (uncertain by more than 1.0 dex) with $\lambda \leq
2.5$ $\mu$m (NIRISS only) transmission spectra for solar composition
exoplanet atmospheres with clouds. Mixing ratio uncertainties improve
somewhat when $\lambda \geq 5$ $\mu$m data (NIRCam or NIRCam+MIRI) are
added in many cases. However, the resulting constraints are considerably
worse than those of the clear solar atmospheres (as good as 0.2 -- 0.5
dex as discussed above). Mixing ratio uncertainties improve to ~$\sim$1
dex for some molecules in the hot Jupiter and warm sub-Neptune cloudy
atmospheres when these longer wavelengths are also included in the
retrievals. The mixing ratio uncertainties of the molecules in the warm
Neptune system also improve with these longer wavelength retrievals, but
none get close to 1.0 dex. The cool super-Earth system retrievals only
produce molecular mixing ratio limits for the cloudy atmosphere case
regardless of wavelength range used.

Cloud-top pressure P$_{\rm cloud}$ is also retrieved for the
transmission cases, and Table~\ref{tbl-4} shows that this is constrained
to $\sim$1 dex for the hot Jupiter and warm sub-Neptune systems when
using complete $\lambda = 1 - 11$ $\mu$m spectra of cloudy solar
atmospheres. This allows locating the position of clouds to an order of
magnitude in pressure in their atmospheres. Reducing the wavelength
range to $1 - 5$ and $1 - 2.5$ $\mu$m increases the uncertainty to
$\sim$1.4 and $\sim$1.7 dex, respectively. The warm Neptune system is
not constrained quite as well (mostly due to its less favorable
transmission parameters), having an uncertainty in P$_{\rm cloud}$ of
1.4 -- 1.9 dex for the 3 different wavelength ranges of the cloudy solar
atmosphere case. The retrievals provide lower limits for P$_{\rm cloud}$
(highest cloud altitudes) for all clear atmospheres (solar or HMMW) of
the hot Jupiter, warm Neptune, and warm sub-Neptune systems. Clouds in
the clear solar atmospheres of all three systems are constrained to P
$\gtrsim$ 1 mbar, a useful limit given that transmission spectra probe
only high altitude / low pressure atmospheric regions. P$_{\rm cloud}$
is not constrained well for any atmospheres of the cool super-Earth
system. The P$_{\rm cloud}$ histograms in
Figures~\ref{fig:HJ_gasses_fig} -- \ref{fig:CSE_gasses_fig} illustrate
these constraints graphically.

\subsection{Emission Spectra Constraints}

Emission spectra retrievals constrain the mixing ratios of the most
dominant species of the solar composition hot Jupiter (better than 0.3
dex for CO, CO$_2$, and H$_2$O) and warm Neptune (to $\sim$1 dex for
CH$_4$, H$_2$O, and NH$_3$) atmospheres. Figures~\ref{fig:HJ_gasses_fig}
- \ref{fig:CSE_gasses_fig} and Table~\ref{tbl-4} show that $\lambda > 5$
$\mu$m data (e.g., MIRI) are required to obtain these good constraints.
This can be understood by examining Fig. \ref{fig:WNTr_emis_spec_fig};
emission spectra have low SNR at shorter wavelengths, and there are
numerous strong molecular absorption features at $\lambda > 2.5$ $\mu$m.
Note that the NIRISS-only ($\lambda = 1 - 2.5$ $\mu$m) emission spectrum
of the hot Jupiter gives a false peak in its CH$_4$ mixing ratio, and
this disappears when longer wavelength data are added (see
Fig.~\ref{fig:HJ_gasses_fig}). Molecular mixing ratio uncertainties are
worse than $\sim$1 dex for the emission retrievals of the warm
sub-Neptune planet, and the SNR of the emission spectrum of the cool
super-Earth was too low to perform useful retrievals at any wavelengths
(see also Fig.~\ref{fig:WNTr_emis_spec_fig}).

These results show that JWST $\lambda > 2.5$ $\mu$m emission spectra
with moderate-to-high SNR will be very useful for atmospheric
characterization. Emission spectra are also required to retrieve T-P
profiles (Figure \ref{fig:TP_summary_fig}), particularly important for
understanding energy absorption and transport in strongly insolated
atmospheres. \citet{F05} showed that the direct, face-on geometry of
emission spectra are much less impacted by clouds or hazes that the
slant geometry of transmission spectroscopy observations. Therefore we
assume that observations and retrievals of emission cloudy solar
atmosphere emission spectra will be very similar to the clear solar
composition atmosphere emission spectra studied here. Given that,
retrievals of JWST $\lambda = 1 - 11$ $\mu$m emission spectra provide
significantly better constraints than the transmission spectra on the
mixing ratios of most molecules for cloudy hot Jupiter and warm Neptune
atmospheres (see Table \ref{tbl-4}). Each star+planet system should be
evaluated to determine whether emission or transmission observations
are more favorable for detecting spectral features and constraining
parameters of interest via atmospheric retrievals. It would
be ideal to acquire and combine both transmission and emission spectra
to probe planetary atmospheres over the broadest
possible pressure range and to use all data to constrain compositions
and chemistries \citep{G14, KBD14b}.

Spectral absorption features are clearly suppressed over the entire
$\lambda = 1 - 11$ $\mu$m wavelength range of the hot Jupiter
temperature inversion emission spectrum
(Fig.~\ref{fig:WNTr_emis_spec_fig}), and the inversion is detected at
modest to high SNR in the retrievals. As shown in Fig.
\ref{fig:TP_summary_fig}, the temperatures of the hot Jupiter
atmosphere decreases to $T = 1260$ K at $P = 10^{-2}$ bar, and then
increases to $T = 2100$ K at $P = 10^{-4}$ bar for the inversion
model. This difference of $\Delta T = 840$ K is detectable in the
retrieved T-P profiles of all 3 wavelength ranges. At these pressures,
the 1 $\sigma$ temperature uncertainties of the $1 - 2.5$, $1 - 5$,
and $1 - 11$ $\mu$m wavelength ranges are approximately 215 K, 40 K,
and 30 K respectively. Therefore the temperature inversion is detected
at better than 4$\sigma$, 21$\sigma$, and 28$\sigma$ respectively when
using multiple retrieved T-P points in this pressure range. It is
clear that the $\lambda > 2.5$ $\mu$m data constrain the hot Jupiter
T-P profile much better than the $\lambda = 1 - 2.5$ $\mu$m data
alone, and Fig. \ref{fig:TP_summary_fig} shows that $\lambda = 1 - 11$
$\mu$m data constrain T-P profiles somewhat better than $\lambda = 1 -
5$ $\mu$m data for all 3 planets studied.

\subsection{HMMW Atmospheres and [Fe/H]}

The compositions of high mean molecular weight (including 100\% H$_2$O)
atmospheres will also be possible to constrain with JWST spectra.
Figures~\ref{fig:WN_gasses_fig} -- \ref{fig:CSE_gasses_fig} show that
the transmission retrievals of the warm Neptune, warm sub-Neptune and
cool super-Earth HMMW atmospheres have significantly different
molecular mixing ratio distributions than the solar ones (with or
without clouds; as expected from Table~\ref{tbl-mol}). Likewise, their
[Fe/H] distributions are also significantly different, reflective of
the high metallicity ([Fe/H] $\simeq$ 3) required to produce HMMW
atmospheres. These differences are readily apparent in the $\lambda = 1
- 2.5$ $\mu$m results alone in most cases, but they become more clear
when longer wavelength data are included. Figures
\ref{fig:WN_gasses_fig} -- \ref{fig:WSN_gasses_fig} and Table
\ref{tbl-4} clearly show that the emission retrievals constrain the
warm Neptune and warm sub-Neptune clear HMMW atmospheres less well than
transmission ones. These figures also show that the derived [Fe/H]
values have uncertainties of $\sim$0.5 dex (factor of 3) or better for
the clear (HMMW and solar composition) atmospheres of the hot and warm
planets studied with $\lambda = 1 - 5+$ $\mu$m transmission spectra.

\subsection{Potential missing species and opacity uncertainties}

Real planets may have atmospheric species not included in the forward
models or retrievals we have performed here. If this were the case,
performing these retrievals on real JWST data would produce
wavelength-dependent residual differences between observed and modeled
spectra. Additional species could then be introduced into the model to
minimize these residuals. Computing the Bayes factor between the models
with simpler and more complex compositions would reveal whether adding
the additional species is statistically justified. Uncertain molecular
opacities may also skew the retrieved compositions for observed
planets. However, this error is likely to be on the order of 10\% but can be as high as a factor of a few in some cases \citep[][\S 3.4]{GH15}. This is mostly less than or equal to our best 68\% mixing ratio confidence widths (see Table \ref{tbl-4}).

\section{Discussion} \label{sec:Discussion}

Here we discuss how applying these precision molecular abundance and
temperature constraints may address several significant outstanding
scientific issues. We focus on questions probing planet formation and
disequilibrium chemistry and end with observational considerations that
will impact data quality and the uncertainties of retrieved values.

\subsection{Carbon-to-Oxygen Ratios}

The C-to-O ratio (C/O) of a planetary atmosphere could be a tracer of
its formation and migration history within a protoplanetary disk
\citep{OMB11}. Determining the C/Os of a diverse set of exoplanets over
a wide range of conditions is important for applying this theory to
understand planet formation scenarios. There has yet to be unambiguous
evidence for high C/O (C/O $>$ 1) within the current ensemble of
observed exoplanets \citep{KBM13, BKM15, MHS11, LKW14, SBM14, KLB15,
B15}. Only upper limits \citep{KLB15, B15} have been derived from
transmission spectra. Emission spectra retrievals have also provided
upper limits for some planets or otherwise unconstrained, and a
specific C/O value has only been determined for HD189733b which has
broad near continuous wavelength coverage from $\sim$1 - 20 $\mu$m
\citep[][]{LKW14, LFI12, WTR15}.

There are several ways of determining or diagnosing the C/O in
exoplanet atmospheres. The most straightforward, direct approach, is to
simply determine the abundance of all of the carbon and oxygen bearing
species present in the spectra and compute it directly by summing the
carbon atoms in all of the carbon species and dividing by the oxygen
atoms in all the oxygen-bearing species. Certainly it is possible for
some carbon or oxygen to be locked away in condensates or other species
that may not necessarily present themselves spectroscopically resulting
in a potential bias. In any case, this is the approach we take. Figures
\ref{fig:HJ_gasses_fig} -- \ref{fig:CSE_gasses_fig} show the C/O
histograms derived from the retrieved molecular abundances. In some
cases, when the abundances are not well constrained, a two peak
distribution exists. This is simply due to the propagation of
uniform-in-log priors through the ratio used to compute the C/O
\citep[see][for an in depth discussion]{LWZ13}. Another approach is to
retrieve the C/O directly as done by \citet{KLB15}, \citet{B15}, and in
\S\ref{sec:Parameterization} for a test case. However doing so
requires the {\it assumption} of the chemical processes (e.g.,
thermochemical equilibrium, vertical mixing, photochemistry) via a
chemical model that relates the elemental abundances to the molecular
abundances (see \S\ref{sec:Parameterization}).
 
We find that for all of the hot Jupiter transmission and emission
scenarios observed with wavelengths longer than 2.5 $\mu$m (e.g.,
NIRISS+NIRCam and NIRISS+NIRCam+MIRI) the C/O can be constrained to
better than 0.2 dex (factor of 1.6) and are largely unbiased by the
two-peak problem. Wavelengths below 2.5 $\mu$m do not capture the
strong CH$_4$, CO, or CO$_2$ vibrational bands that occur at longer
wavelengths. The low abundance of CH$_4$ requires observation of the
strong 3.3 and / or 7.7 $\mu$m band to provide a meaningful upper
limit as CH$_4$ is not thermochemically dominant in hot atmospheres.
While CO and CO$_2$ are dominant in hot atmospheres, they have a
relatively narrow influence across wavelength. Furthermore, there is
little improvement extending coverage beyond 5 $\mu$m as the strongest
CO or CO$_2$ bands over the full $1-11$ $\mu$m wavelength range occurs
at $\sim$4.5 $\mu$m (see Figure 1). Therefore, for hot Jupiters with
comparable SNRs, observations over $\lambda = 1-5$ $\mu$m are more
than sufficient to provide a meaningful constraint on their C/O.

The transmission spectra of warm Neptune and sub-Neptune low mean molecular weight (solar composition) clear atmospheres also unbiasedly constrain the C/O to $\sim$0.2 dex or better in all three wavelength ranges. This is because the CH$_4$ is largely constrained to abundances that are greater than the CO or CO$_2$ abundances and thus the C/O ratio is mainly set by the methane-to-water ratio (CH$_4$/H$_2$O). Because of the large abundance of CH$_4$ in the cooler planets, it presents strong spectral features at wavelengths below 2.5 $\mu$m (see Figure~\ref{fig:gas_perturbations_spec_fig}).  Even in the cloudy scenarios the log(C/O) constraints are only a factor of 1.2 worse. 

Emission spectra constrain C/O less well in these small warm planets.
$1-11$ $\mu$m emission spectra constrain C/O to $\sim$0.5 dex in the
warm Neptune solar composition atmosphere, and $1-5$ $\mu$m emission
spectra also provide some constraint (to only $\sim$1 dex). The
$\lambda \geq 5$ $\mu$m emission spectra provide only loose
constraints on C/O in the warm sub-Neptune clear solar atmosphere.
Even complete $1-11$ $\mu$m emission spectra produce C/O distributions
very similar to the priors for the high mean molecular weight
atmospheres of these planets. These relatively cool planets have low
flux and therefore low SNR near the $\lambda \leq 4$ $\mu$m CH$_4$
bands, so these results are not particularly surprising.

In general, the C/O constraints are poor for the cool super-Earth atmospheres. None of the transmission scenarios constrain C/O to significantly better than an order of magnitude.

\begin{figure} % figure* spans 2 columns
\centering
\includegraphics[width=3.5in]{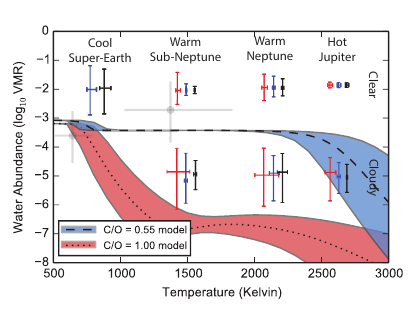}
%\hspace{0.5 in}%
\caption{
Application of precision H$_2$O and scale height temperature measurements from transmission spectra to distinguishing between high and low C/O scenarios \citep[after][]{KLB15}.  The broad red and blue curves are equilibrium chemistry models that show how the H$_2$O abundance changes as a function of temperature for high (red) and low (blue) C-to-O ratios for solar metallicity.  The spread in the equilibrium chemistry models is due to different assumptions regarding the probed pressure levels (0.1 - 10 mbar). We show representative error bars derived from our temperature and H$_2$O retrieval results for the solar metallicity scenarios observed in transmission (Table~\ref{tbl-4}).  The red error bars represent the NIRISS only, blue are for NIRISS+NIRCam, and black are for NIRISS+NIRCam+MIRI.   The top set error symbols show the constraints in the clear atmospheres and the bottom show cloudy ones.  {\em These positions are located for clarity and are not indicative of actual retrieved temperatures or compositions!} The light gray points and error bars are constraints from HST WFC3 transmission measurements of WASP-12b  \citep{KLB15} (hotter) and WASP-43b {KBD14b} (cooler). 
}
\label{fig:H2OvT} 
\end{figure}

Finally, Figure \ref{fig:H2OvT} shows how the retrieved H$_2$O mixing
ratio uncertainties compare to the predicted thermochemical H$_2$O
values for two different C/O cases \citep[see][]{KLB15}. The
uncertainties in H$_2$O mixing ratios are shown for the different
planets' transmission cases (their positions are located for clarity
and are not indicative of true temperatures or compositions). The
H$_2$O mixing ratio depends on metallicity and contains no direct
information on carbon abundance, so it cannot provide an unambiguous
C/O constraint by itself. However it is instructive to show how one
could use H$_2$O to rule out a high C/O scenario in some cases.
For hot Jupiters, where the difference between solar and high C/O
chemical models is largest, a measurement in a clear atmosphere could
distinguish the difference between a solar and high C/O atmosphere by
greater than 50$\sigma$ for all 3 wavelength ranges. Using H$_2$O alone
becomes difficult for planets with scale height temperatures below
$T \sim$1000 K as the predicted thermochemical H$_2$O abundances
converge for both solar value and high C/O. Other proxies such as the
ratio of H$_2$O to CH$_4$ \citep[under the assumption that oxidized
carbon will be in low abundance; see Fig. 2 of ][]{M12} or a direct
determination of the C/O (as above) will be required.

Regardless of the methodology used, it appears that JWST spectra will
be very useful in determining the C-to-O ratios in a wide array of
planetary atmospheres and will allow us to begin to address
quantitatively the formation location relative to ice lines and any
subsequent migration.

%%%%%%%%%%%%% figure %%%%%%%%%%%%%%%%%%%%
%%%%%%%%%%%% figure %%%%%%%%%%%%%%%%%%%%

\subsection{The Mass-Metallicity Relationship}

Understanding planet formation requires determining and interpreting the
mass-metallicity relationship for a varied sample of planets.
Establishing this relationship over the exoplanet population provides
insight into the core accretion formation mechanism
\citep[e.g.,][]{IL05}. There has been some progress on establishing the
role that planetary mass and parent star metallicity have in determining
the \emph{bulk} metallicity of transiting gas giants \citep{MF11}
through the use of planetary evolution models. A corresponding
constraint on planetary atmospheric metallicity, from spectroscopy,
would help us to understand if most of these metals are found in a core,
or are mixed within the H/He-dominated envelope.

\begin{figure} % figure* spans 2 columns
\centering
\includegraphics[width=3.5in]{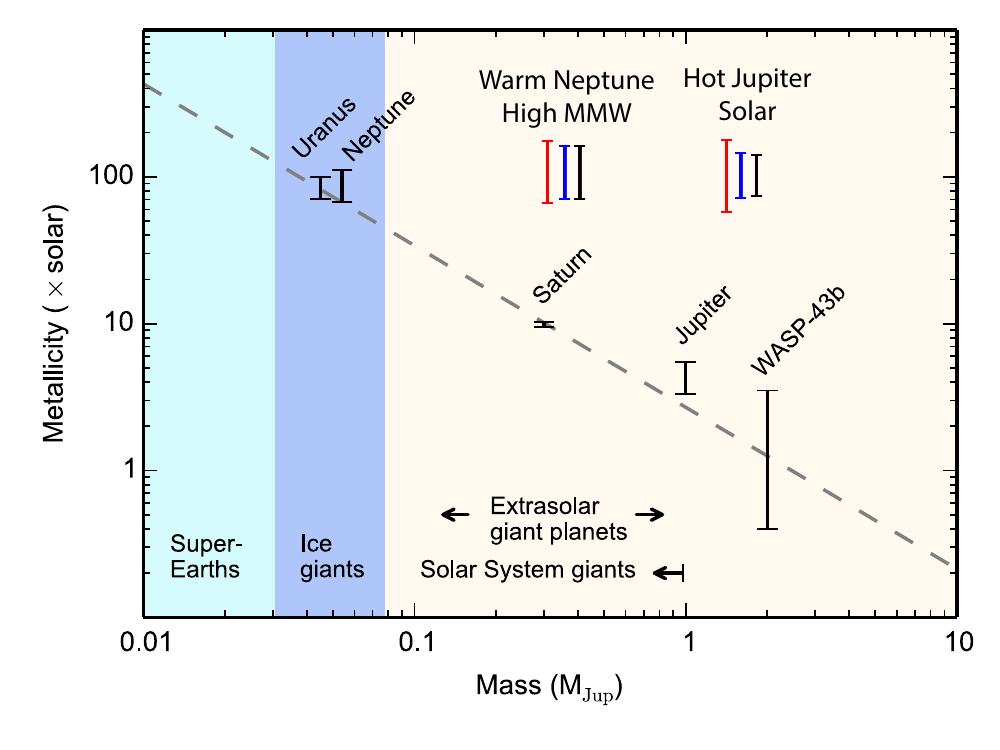}%
\hspace{0.5 in}%
\caption{Atmospheric mass-metallicity relationship (after \citet{KBD14b}).  Solar System
planets and a measured exoplanet are shown as the black points with error
bars.  Representative transmission spectra metallicity constraints for a clear solar composition hot Jupiter and a clear high mean molecular weight warm Neptune are shown near the top ({\it not at their actual mass/metallicity values}). Three 68\% confidence constraints are shown for each: black for NIRISS+NIRCam+MIRI, blue for NIRISS+NIRCam, and red for NIRISS only.
 }
\label{fig:mass_metallicity}
\end{figure}

Using population synthesis models \citep{MAK12}, \citet{FMN13}
suggests that as the mass of a planet decreases, the atmospheric
metallicity increases. Lower mass planets are unable to accrete
substantial envelopes within the core-accretion theory and thus are
more susceptible to pollution by in falling planetesimals. We do
indeed find tantalizing evidence for such a trend within our own solar
system as seen in Figure~\ref{fig:mass_metallicity}. However, in our
own solar system we are limited to using CH$_4$ as the proxy for
metallicity as water is largely sequestered in deep clouds. Exoplanet
atmospheres, due to their high temperatures, permit us to access a
wide array of molecules allowing for more precise constraints on the
envelope metallicity. \citet{KBD14b} provided additional leverage on
this relationship via a relatively good constraint on the water
abundance in a 2 M$_{J}$ hot Jupiter. Under the assumption of solar
C/O, the retrieved water abundance was used as a proxy for
metallicity. This proxy-metallicity was found to be consistent with
the observed solar system trend. Furthermore, constraints of Neptune
mass objects (GJ 436b by \citealt{SHN10} and HAT-P-11b by
\citealt{FDB14} are also suggestive of this trend.

Recently, \citet{B15} demonstrated that the water abundance alone
cannot constrain the atmospheric metallicity because of the degeneracy
of metallicity with the carbon-to-oxygen ratio. The broad wavelength
coverage and high SNR of JWST data enable us to constrain not only the
water abundance, but carbon species as well. This in essence, breaks
the C/O-metallicity degeneracy. We directly compute the metallicity
([Fe/H]) from the retrieved molecular mixing ratios, and the
metallicity histograms are shown in Figures~\ref{fig:HJ_gasses_fig} --
\ref{fig:CSE_gasses_fig} (see \S\ref{sec:Results}). Figure
~\ref{fig:mass_metallicity} compares a typical metallicity constraint
for a solar composition hot Jupiter and a high mean molecular weight
warm Neptune. Observing just 5 planets logarithmically spaced between
a few Jupiter masses and a Neptune mass with such constraints (0.4
dex) would allow us to determine the mass-metallicity slope (in log
space) with a 1$\sigma$ uncertainty of 0.13.

%%%%%%%%%%%%% figure %%%%%%%%%%%%%%%%%%%%
%%%%%%%%%%%% figure %%%%%%%%%%%%%%%%%%%%

\subsection{Disequilibrium Chemistry} \label{sec:Disequilibrium}

Disequilibrium processes are likely to play a role in sculpting the molecular abundances in exoplanet atmospheres \citep{LPL03, ZMF09, LLY10, MVF11, LVC11, VHA12, APV14, HSB13}. These processes come in a variety of flavors including, but not limited to vertical and horizontal mixing \citep{PB77, CS06}, photochemistry \citep{YD99}, ion chemistry \citep{LCV08}, and biology \citep[e.g.,][]{H94}.  The predicted dominant disequilibrium process in jovian-type planets is due to vertical mixing and thus we focus on the ability of JWST to infer disequilibrium due to vertical mixing.  Vertical mixing tends to set (or quench) the upper atmospheric abundance of certain molecules to a particular deep atmospheric value which can result in an orders-of-magnitude enhancement (or depletion) relative to the expected equilibrium abundance at a higher altitude in the atmosphere.  \citet{LY13} devised a scheme to determine the degree to which an atmosphere is out of equilibrium given the retrieved abundances of H$_2$O, CH$_4$, CO, and H$_2$.   If these values form a ratio, $\alpha$, 
\begin{equation}\label{alpha}
\alpha(f_{i},P)=\frac{f_{CH_4}f_{H_2O}}{f_{CO}f_{H_2}^3P^2}=K_{eq}(T)
\end{equation}
where $f_{i}$ are the molecular mixing ratios of species $i$, $P$ is the pressure in bars, and $T$, temperature in K, that is equivalent to the thermochemical equilibrium constant $K_{eq}(T)$, than those species are considered to be in chemical equilibrium and thus disequilibrium mechanisms are weak or non-existent. 

\begin{figure} %[h] % figure* spans 1 column
\centering
\includegraphics[width=3.5in]{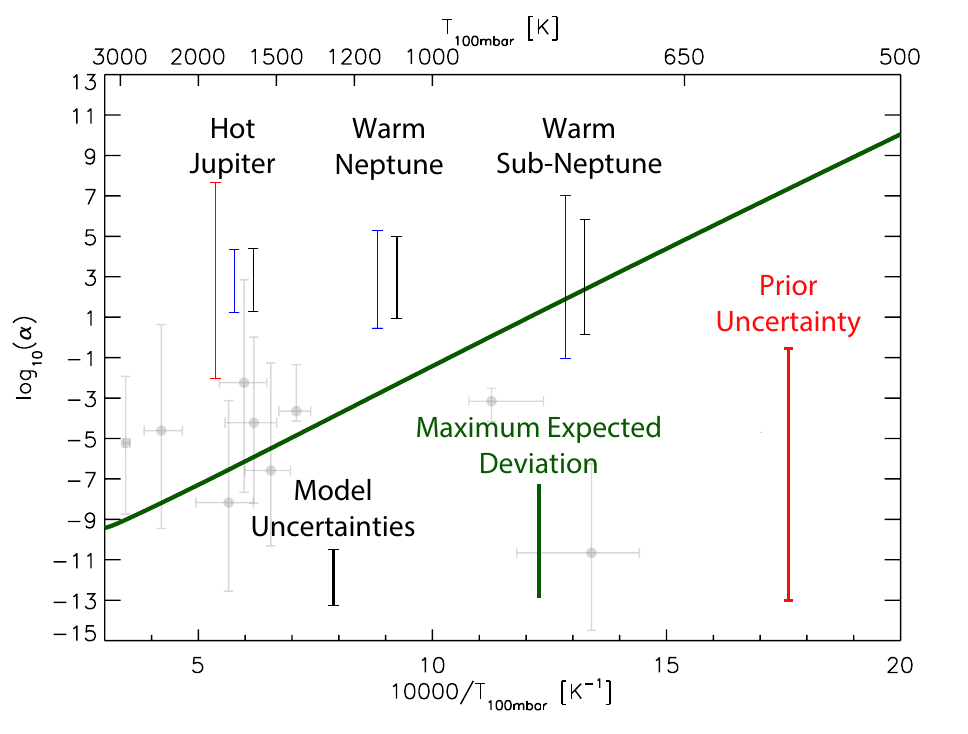}
\caption{   
Diagnosing disequilbirum chemistry with JWST \citep[adapted from][]{LY13}. The green line is the equilibrium constant as a function of
temperature. Given the equilibrium molecular abundances of H$_2$O, CH$_4$, CO and H$_2$ at a given temperature and pressure, the quantity $\alpha$ (see Eq. \ref{alpha}) will fall on this line. If there are strong disequilibrium processes, $\alpha$ will deviate from the green line. $\alpha$ computed from the retrieved abundances for H$_2$O, CH$_4$,
CO and H$_2$ for a variety of planets observed with HST and Spitzer are
shown in light gray \citep{LY13, LKW14}. We show
representative error bars for $\alpha$ derived from the retrieved mixing
ratios of emission spectra for the hot Jupiter, warm Neptune, and warm Sub-Neptune solar
composition emission scenarios for NIRISS only (red), NIRISS+NIRCam (blue), and NIRISS+NIRCam+MIRI (black). {\it Their positions in the plot are arbitrary.} Model uncertainty due to the uncertainty in the pressure levels probed (e.g., the width of the thermal emission contribution function across wavelength) is shown as the black error bar. The maximum expected deviation due to vertical mixing is shown as the green error bar. Finally, the prior uncertainty  is shown as the red curve.
%, that is the uncertainty on $\alpha$ we would derive given the %uniform-in-log priors for each of the gases
  }
\label{fig:Diseq} 
\end{figure}

Figure \ref{fig:Diseq} summarizes the constraints on $\alpha$ as provided by the different wavelength regions for emission sectra.  We also note, as in \citet{LY13}, that there could be up to a factor of $\sim$100 more uncertainty due to the spread in the probed pressure levels.  Furthermore, in some cases only an upper limit on one of the three retrieved gases can be obtained. In these cases $\alpha$ would also only have an upper or lower limit.  We find that the hot Jupiter scenarios will have the best constraint on $\alpha$. Unfortunately hot Jupiters are generally predicted to be in equilibrium \citep{LY13}, so the small error bar is not very useful for identifying disequilibrium in those planets.  The objects cooler than $\sim$1000 K will likely show signs of disequilibrium due to the dredging up of CO (and also N$_2$ in the NH$_3$--N$_2$ chemical system). This will result in an $\alpha$ that falls below the equilibrium line.  The maximum expected deviation due to strong vertical mixing is only $\sim$2$\sigma$ larger than the smallest warm Neptune error bar, making the definitive detection of disequilibrium due to vertical mixing difficult.  Perhaps the best way of approaching this problem is to determine $\alpha$ over a wide range of planetary effective temperatures to identify where deviations from equilibrium begin to occur. This transition is likely to occur between 1000--1200 K \citep[e.g.,][]{MVF11}. This temperature region is also a good place to observe because the CH$_4$ and CO abundances begin to thermochemically trade places. Their near equal abundances and the higher temperatures (relative to the warm Neptunes) will likely allow bounded constraints on both the CO and CH$_4$ abundances providing an actual bounded constraint on $\alpha$. 

We have broadly explored the detectability of disequilibrium due to vertical mixing.  Certainly, photochemistry can produce species not included in this investigation (e.g., C$_{m}$H$_{n}$, HCN), especially in cooler planets in which CH$_4$ is readily available for photolysis \citep{LVC11, VAS14, MLV13, HS14, MZF12}. \citet{SFG11} determined that some of these photoproducts are potentially detectable in JWST observations of planets similar to the warm Neptune (GJ 436b) we have considered here.
   
%%%%%%%%%%%%% figure %%%%%%%%%%%%%%%%%%%%
%%%%%%%%%%%% figure %%%%%%%%%%%%%%%%%%%%

\subsection{Additional Observational Considerations}

The precision of the retrieved parameters are limited by the particular
star-planet systems chosen as well as the planets themselves. For
example, the H$_2$O mixing ratio uncertainty using all wavelengths is
considerably better for the warm sub-Neptune (0.25 dex) versus the warm
Neptune (0.59 dex) clear solar atmosphere in transmission. These two
planets have similar equilibrium temperatures and scale heights (see
Table~\ref{tbl-2}), so this difference is likely due to the higher SNR
of the warm sub-Neptune simulations caused by the relatively high
brightness and small size of the adopted host star, GJ 1214. Each
star+planet system should be evaluated to determine whether emission or
transmission observations are more favorable for detecting spectral
features and constraining parameters of interest. The planetary system
parameters given in Table~\ref{tbl-2} should be useful for scaling these
results to other systems.

We do not know how well co-adding observations of multiple transits or
secondary eclipses will improve our results at this time. The
simulated single transit and single eclipse observations of our
selected systems typically have total noise values only $10 - 50$\%
larger than our adopted noise floors (\S\ref{sec:Sims}), so systematic
noise assumptions have already significantly influenced the precision
of our simulated data and retrieved information. Given this, co-adding
more data would not substantially improve the results for these very
observationally favored systems with bright host stars. We believe
that our systematic noise assumptions are reasonable, but we will not
know their exact values in different conditions (e.g., co-adding
multiple spectra vs. binning to lower resolution) until after
JWST becomes operational.

Other instrumental and astrophysical noise sources may also impact JWST
data. \cite{BAI15} considered the impact of systematic
wavelength-invariant offsets between the spectra acquired in different
JWST instrument modes. They found that the impact on retrieved
parameters was minimal if even small regions of overlap between the
modes were used to correct such errors to below the noise of their
simulations. This is also likely to apply in our study if residual
offsets between the different spectral regions are (corrected to) less
than their respective adopted systematic noise floors. We note that the
different instrument modes have more spectral overlap than the regions
we selected in Table~\ref{tbl-3}.

\cite{BAI15} also note that star spots can impact transmission (but
not emission) spectra considerably \citep[see also][]{PSG13} and
assess their impact on retrievals of planet parameters for both solar-
and M-type ($T_{eff} = 3000$ K) host stars. They find that the quality
of retrieved information is not seriously degraded for a hot Jupiter
with a Sun-like star that has a relatively high spot fraction (3\% of
its area). However, \cite{BAI15} find that the retrieved H$_2$O mixing
ratio can be up to an order of magnitude too high for the transmission
spectrum of a hot Neptune planet with an M-type host star with a spot
coverage of 10\% if the planet does not occult any spots during
transit. This error is considerably larger than the H$_2$O mixing
ratio uncertainty we expect for the warm Neptune with a clear
solar composition atmosphere and full $1 - 11$ $\mu$m wavelength
coverage (Table~\ref{tbl-4}), and we share the concern with
\cite{BAI15} that stellar activity could be the limiting factor for
accurate retrievals of JWST observations of active stars. Simultaneous
photometric observations at relatively short wavelengths \citep[e.g.,
see][]{FDB14} could be used to correct this effect, and this may be
possible with NIRCam $\lambda \leq 2.4$ $\mu$m imaging during the NIRCam
$\lambda = 2.5 - 5.0$ $\mu$m grism spectral observations.

\section{Summary and Conclusions} \label{sec:Summary}

We have generated forward models of transmission and emission spectra
of generic hot Jupiter, warm Neptune, warm sub-Neptune, and cool
super-Earth planets using the parameters of well-known systems. We
then performed simulations of slitless JWST observations using the
instrument modes NIRISS SOSS over $1 - 2.5$ $\mu$m, the NIRCam grisms
over $2.5 - 5.0$ $\mu$m (2 exposures required), and the MIRI LRS over
$5.0 - 11$ $\mu$m. The NIRSpec instrument may be used instead of
NIRCam, but it also requires 2 exposures over $2.5 - 5.0$ $\mu$m for
bright stars, and we understand its likely systematic noise less well
than NIRCam at this time. We performed retrievals on single simulated
transits and secondary eclipses to assess what information and
constraints JWST data are likely to provide on these planet archetypes
and other exoplanet atmospheres. We arrive at following major
conclusions:

\begin{enumerate}

\item JWST will likely obtain high quality transmission and emission
spectra of a variety of exoplanet atmospheres over a wavelength range
of at least $1 - 11$ $\mu$m. Obtaining spectra over this entire range
will typically require observations of 4 separate transit or secondary
eclipse events using 4 instrument modes for planets with bright host
stars. We do not know at this time exactly how JWST data will be
impacted by systematic noise, but we have assumed noise floors
consistent with the best performance of HST or Spitzer as appropriate
for each wavelength range. 
	
\item The volume mixing ratios of dominant molecular species, C/O, and
[Fe/H] of clear solar composition planetary atmospheres can be
diagnosed very well with transmission spectra, often with only $1 -
2.5$ $\mu$m wavelength NIRISS data. However, longer wavelength data are
needed for good constraints in some cases. Strong temperature
inversions should be detectable in the T-P profile of the clear solar
composition hot Jupiter planet with emission spectra covering $\lambda
= 1 - 2.5$ $\mu$m. The T-P profiles of the hot Jupiter, warm Neptune,
and warm sub-Neptune are all constrained reasonably well with $1 - 5$
$\mu$m emission spectra. JWST transmission spectra will be useful for
discriminating between cloudy and HMMW atmospheres of small planets in
observationally favorable systems.

\item JWST spectra have the potential to constrain the molecular mixing
ratios, C/O, [Fe/H], and T-P profiles of planets with cloudy
atmospheres. Mixing ratios of the cloudy warm sub-Neptune is constrained
to $\sim 1$ dex with transmission spectra covering $\lambda = 1 - 11$
$\mu$m. However, $\lambda = 1 - 11$ $\mu$m emission spectra provide
significantly better constraints than the transmission spectra on the
mixing ratios of most molecules for cloudy hot Jupiter and warm Neptune
atmospheres assuming that emission spectra are not impacted by clouds.
Each star+planet system should be evaluated to determine whether
emission or transmission observations are more favorable for detecting
spectral features and constraining parameters of interest. Emission
spectra can be more useful than transmission in cases of sufficiently
high $F_p$ accompanied by high $F_p/F_*$. Acquiring both transmission
and emission spectra will probe exoplanet atmospheres over the broadest
possible pressure ranges and constrain compositions and chemistries
better than either transmission or emission alone.

\item The complete $\lambda = 1 - 11$ $\mu$m cool super-Earth
emission spectrum had insufficient signal-to-noise for retrievals. There
simply is not enough flux contrast $F_p/F_*$ for useful emission spectra
from this system when a single secondary eclipse is observed at each
wavelength. Photometric filter observations may be more useful for
constraining the planet's properties. Small ($R\lesssim 2 R_\oplus$),
cool ($T<$700 K) planets will need host stars with $K \lesssim 8.5$ mag
and / or spectral types later than M0 V for useful emission spectra of
single secondary eclipses.

\item  The molecular mixing ratios retrieved from  $\lambda = 1 - 5+$ $\mu$m JWST transmission spectra will provide derived [Fe/H] values with an uncertainty of 0.5 dex (factor of 3) or better for the clear atmospheres of the hot and warm planets studied. This is adequate for evaluating systematic differences in metallicity with planet mass and determining whether this function is substantively similar to or different from our Solar System.

\item Carbon-to-oxygen ratios derived from the retrieved molecular
mixing ratios are constrained to better than 0.2 dex (factor of 1.6)
for the hot Jupiter system when using $\lambda \geq 5$ $\mu$m
transmission or emission spectra. H$_2$O mixing ratio values and
retrieved atmospheric temperatures can also be used to distinguish
between high (1.0) and solar (0.55) C/O values for hot planets (and
with only $\lambda \leq 2.5$ $\mu$m transmission data in many cases),
sufficient for assessing whether they formed interior to or exterior
of the H$_2$O ice lines in their protoplanetary disks. Transmission spectra constrain C/O better than emission spectra for the warm Neptune and sub-Neptune systems. $\lambda \geq 5$ $\mu$m spectra provide the best constraints for these warm planets, but $\lambda = 1 - 2.5$ $\mu$m data give good results for clear solar atmospheres. 

\item The uncertainties in molecular mixing ratios retrieved from JWST emission spectra (even $\lambda \geq 5$ $\mu$m) for single observations are too large to obtain a definitive detection of vertical mixing via the method of \citet{LKD13}. However, observing many planets that span a range of effective temperatures should permit us to identify at which temperatures the molecular abundances deviate from equilibrium. The ensemble of retrieved mixing ratio values and uncertainties of the planets studied here can be used to assess how well each species can be detected in observations of real planets that may have different compositions (e.g., the sensitivity to CO or CO$_2$ in cool planets). The detection limits presented here should be scaled to the SNR of any real data, and similar wavelength ranges should be used in such comparisons.

\item We not know at this time how well SNR and retrieval uncertainties
will improve with the binning or co-addition of data beyond the single
transit or eclipse observations simulated here. We hope and expect that
this work will be useful for planning early JWST observations, but
actual on-orbit performance must be measured to make better predictions.

\end{enumerate}

\acknowledgments

We are grateful to L. Albert, J. Barstow, J. Bean, S. Birkmann,
J. Bouwman, R. Doyon, P. Ferruit, Th. Henning, L. Kreidberg,
P.-O. Lagage, N. Lewis, M. Marley, and the JWST NIRCam and MIRI
instrument teams for helpful science discussions, feedback, and
information on instrument performance. We also thank the referee I.
Crossfield whose numerous insightful recommendations and comments
allowed us to substantially improve the paper. This research has made
use of the Exoplanet Orbit Database and the Exoplanet Data Explorer at
exoplanets.org. The simulations for this research were carried out on
the UCSC supercomputer Hyades, which is supported by National Science
Foundation (award number AST-1229745) and University of California,
Santa Cruz. TPG acknowledges support from the NASA JWST Project and
Program for this work via WBSs 411672.04.01.02 and 411672.05.05.02.02.
MRL acknowledges support provided by NASA through Hubble Fellowship
grant \#51362 awarded by the Space Telescope Science Institute, which
is operated by the Association of Universities for Research in
Astronomy, Inc., for NASA, under the contract NAS 5-26555. JJF
acknowledges the support of NSF grant AST-1312545.

{\it Facilities:} \facility{JWST}.

%\newpage

%% empulateapj.sty: from http://hea-www.harvard.edu/~alexey/emulateapj/emulateapj.cls
%% 3) Multi-page tables cannot be set properly inside the main text; you
%% need to move the table to the end of the paper (after the references) and
%% issue the command \LongTables before it. 

%\clearpage
%\LongTables
\begin{deluxetable*}{lllrrrrrrrrrr}
\tabletypesize{\scriptsize}
%\rotate
\tablecaption{Retrieved Parameter Uncertainties or Limits\tablenotemark{a} \label{tbl-4}}
\tablewidth{0pt}
\tablehead{
\colhead{Planet} & \colhead{$\lambda$ ($\mu$m)} & \colhead{Geom} &
\colhead{T(K)} & 
\colhead{xRp} & \colhead{log(P$_{c}$) (bar)} &
\colhead{logH$_2$O} & \colhead{logCH$_4$} & \colhead{logCO} & \colhead{logCO$_2$} & \colhead{logNH3} & \colhead{logN2} 
}
\startdata
\\
\multicolumn{2}{l}{\bf{Hot Jupiter}}\\ 
Clear Solar & $1 - 11$ & Trans & 25 & 0.002 & $>$-0.82 & 0.16 & $<$-7.1 & 0.32 & 0.23 & $<$-7.1 & $<$-2.7 \\
 & $1-5.0$ & Trans & 28 & 0.002 & $>$-0.77 & 0.17 & $<$-7.1 & 0.34 & 0.25 & $<$-6.5 & $<$-2.7 \\
 & $1-2.5$ & Trans & 35 & 0.002 & $>$-0.82 & 0.20 & $<$-6.0 & 1.8 & $<$-5.1 & $<$-6.0 & $<$-2.5 \\ 
 &  &  &  &  &  &  &  &  &  &  &  & \\ 
Cloud Solar & $1 - 11$ & Trans & 18 & 0.01 & 1.1 & 1.1 & $<$-5.6 & 1.3 & 2.9\tablenotemark{b} & $<$-5.5 & $<$-2.4 \\
 & $1-5.0$ & Trans & 33 & 0.011 & 1.3 & 1.5 & $<$-5.4 & 1.6 & 3.3\tablenotemark{b} & $<$-4.6 & $<$-1.8 \\ 
 & $1-2.5$ & Trans & 180 & 0.021 & 1.6 & 1.5 & $<$-3.3 & $<$-1.3 & $<$-2.6 & $<$-3.6& $<$-0.80 \\ 
 &  &  &  &  &  &  &  &  &  &  &  & \\ 
Solar & $1 - 11$ & Emis & &  &  & 0.17 & $<$-6.9 & 0.26 & 0.17 & $<$-6.4 & \nodata \\ 
 & $1-5.0$  & Emis & \nodata &  \nodata &  \nodata & 0.22 & $<$-6.9 & 0.36 & 0.23 & $<$-5.9 & \nodata  \\ 
 & $1-2.5$ & Emis & \nodata &  \nodata &  \nodata & 0.63 & $<$-4.6 & $<$-1.1 & $<$-4.6 & $<$-5.1 & \nodata  \\ 
\\
Solar & $1 - 11$ & Emis & \nodata &  \nodata &  \nodata & 0.32 & $<$-6.1 & 0.38 & 0.41 & $<$-5.5  & \nodata \\ 
 Inversion & $1-5.0$ &  Emis & \nodata &  \nodata &  \nodata & 0.36 & $<$-5.8 & 0.44 & 0.53 & $<$-5.2 & \nodata  \\ 
 & $1-2.5$  &  Emis & \nodata &  \nodata &  \nodata & 3.6 & $<$-2.5 & $<$-2.5 & $<$-2.5 & $<$-5.0 & \nodata  \\ 
\hline
\\
\multicolumn{2}{l}{\bf{Warm Neptune}}\\ 
Clear Solar & $1 - 11$ & Trans & 17 & 0.008 & $>$-1.3 & 0.59 & 0.51 & $<$-3.5 & $<$-7.1 & 0.53 & $<$-2.5 \\ 
 & $1-5.0$  & Trans & 19 & 0.010 & $>$-1.4 & 0.71 & 0.61 & $<$-3.3 & $<$-7.1 & 0.63 & $<$-2.4 \\ 
 & $1-2.5$ & Trans & 42 & 0.014 & $>$-1.7 & 0.91 & 0.87 & $<$-2.3 & $<$-3.7 & 0.85 & $<$-2.3 \\ 
\\
Cloud Solar & $1 - 11$ & Trans & 76 & 0.025 & 1.6 & 1.7 & 1.7 & $<$-3.1 & $<$-5.0 & 1.7 & $<$-1.2 \\ 
 & $1-5.0$  & Trans & 74 & 0.020 & 1.5 & 1.6 & 1.5 & $<$-2.9 & $<$-4.9 & 3.6\tablenotemark{b} & $<$-1.7 \\ 
 & $1-2.5$  & Trans & 180 & 0.029 & 1.9 & 2.0 & 1.9 & $<$-0.9 & $<$-2.4 & $<$-1.8 & $<$-0.50 \\ 
\\
 High MMW & $1 - 11$ & Trans & 170 & 0.005 & $>$-3.2 & 3.9\tablenotemark{b}  & 1.6 & $<$-0.10 & 1.6 & $<$-2.5 & $<$-0.20 \\ 
 & $1-5.0$ & Trans & 240 & 0.004 & $>$-2.8 & 3.7\tablenotemark{b} & 1.8 & $<$-0.20 & 1.8 & $<$-2.7 & $<$-0.20 \\ 
 & $1-2.5$  & Trans & 300 & 0.005 & $>$-3.4 & $<$-0.06 & $>$-4.0 & $<$ -0.20 & 2.7 & $<$-1.9 & $<$-0.20 \\ 
\\
Solar & $1 - 11$ &  Emis & \nodata &  \nodata &  \nodata & 1.26 & 0.79 & $<$-3.6 & $<$-6.1 & 1.2 & \nodata  \\ 
 & $1-5.0$  &  Emis & \nodata &  \nodata &  \nodata & 2.4 & 1.5 & $<$-2.4 & $<$-5.7 & 4.7\tablenotemark{b} & \nodata  \\ 
\\
High MMW & $1 - 11$ &  Emis & \nodata &  \nodata &  \nodata & $<$-1.6 & 2.1 & $<$-1.2 & $>$-6.2 & $<$4.1 & \nodata \\ 
 & $1-5.0$  &  Emis & \nodata &  \nodata &  \nodata & -- & 4.3 & $<$-1.8 & -- & $<$-0.7 &  \\ 
\hline
\\
\multicolumn{2}{l}{\bf{Warm Sub-Neptune}}\\
Clear Solar & $1 - 11$ & Trans & 8.1 & 0.006 & $>$-0.70 & 0.25 & 0.19 & $<$-6.4 & $<$-8.0 & 0.21 & $<$-2.0 \\ 
 & $1-5.0$ & Trans & 9.8 & 0.009 & $>$ -0.8& 0.40 & 0.30 & $<$-6.1 & $<$-7.9 & 0.31 & $<$-2.5 \\ 
 & $1-2.5$ & Trans & 33 & 0.029 & $>$-1.6 & 1.1 & 1.0 & $<$-2.4 & $<$-3.7 & 1.0 & $<$-2.1 \\ 
\\
Cloud Solar & $1 - 11$ & Trans & 22 & 0.024 & 0.92 & 1.0 & 0.96 & $<$-4.0 & $<$-5.7 & 1.0 & $<$-2.4 \\ 
 & $1-5.0$  & Trans & 25 & 0.044 & 1.6 & 1.7 & 1.6 & $<$-5.4 & $<$-5.9 & 1.6 & $<$-2.2 \\ 
 & $1-2.5$  & Trans & 170 & 0.076 & 1.9 & 2.1 & 2.1 & $<$-1.2 & $<$-6.4 & 2.4 & $<$-0.80 \\ 
\\
High MMW & $1 - 11$ & Trans & 100 & 0.004 & $>$-3.3 & 1.0 & 0.65 & $<$-0.4 & 0.84 & $<$-2.4 & $<$-0.10 \\ 
  & $1-5.0$  & Trans & 130 & 0.005 & $>$-3.2 & 1.4 & 1.1 & $<$-0.10 & 1.4 & $<$-2.6 & $<$-0.10 \\ 
 & $1-2.5$  & Trans & 200 & 0.007 & $>$-3.4 & $>$-5.0 & 1.2 & $<$-0.20 & $<$-0.50 & $<$-1.8 & $<$-0.10 \\ 
\\
Solar & $1 - 11$ & Emis & \nodata &  \nodata &  \nodata & 2.2 & 1.5 & $<$-1.4 & $<$-4.7 & 2.2 & \nodata \\ 
 & $1-5.0$  & Emis & \nodata &  \nodata &  \nodata & 6.1\tablenotemark{b} & 2.5 & $<$-1.7 & $<$-4.6 & $<$-0.50 & \nodata  \\ 
\\
 High MMW & $1 - 11$ &  Emis & \nodata &  \nodata &  \nodata & 4.3\tablenotemark{b} & 2.3 & $<$-0.50 & 3.2\tablenotemark{b} & $<$-0.80 & \nodata  \\ 
 & $1-5.0$  &  Emis & \nodata &  \nodata &  \nodata & $<$-0.10 & -- & $<$ -2.1 & 5.0\tablenotemark{b} & -- & \nodata  \\ 
\hline
\\
%\tablebreak
\multicolumn{2}{l}{\bf{Cool Super-Earth}}\\ 
Clear Solar & $1 - 11$ & Trans & 72 & 0.030 & $>$-2.4 & 1.8 & 1.3 & $<$-2.1 & $<$-3.4 & $1.3$ & $<$-1.1 \\ 
 & $1-5.0$ & Trans & 85 & 0.028 & $>$-2.5 & 1.9 & 1.3 & $<$-2.1 & $<$-3.3 & 1.3 & $<$-1.1 \\ 
 & $1 - 2.5$ & Trans & 120 & 0.035 & $>$-2.8 & $<-0.7$ & 1.4 & $<$-1.4 & $<$-1.9 & 1.4 & $<$-0.20 \\
\\
Cloud Solar & $1 - 11$ & Trans & 570 & 0.25 & 2.8 & $<$-0.10 & $>$-7.5 & $<$-0.11 & $<$-0.44 & -- & -- \\ 
 & $1-5.0$  & Trans & 700 & 0.200 & $>$1.5 & $<$-0.10 & $>$-7.5 & $<$-0.28 & $<$-0.27 & -- & -- \\ 
 & $1 - 2.5$  & Trans & $<$3000 & 0.33 & $>$1.5 & -- & -- & -- & --& -- & -- \\ 
\\
Clear H$_2$O & $1 - 11$ & Trans & 670 & 0.043 & $>$-5.9 & $>$-5.0 & $<$-1.8 & -- & $<$-0.10 & $<$-0.14& --\\ 
 & $1-5.0$  & Trans & 720 & 0.044 & $>$-5.9 & $>$-7.0 & $<$-1.4 & -- & $<$-0.10 & $<$-0.50 & -- \\ 
 & $1 - 2.5$ & Trans & 970 & 0.075 & $>$-5.9 & $>$-7.0 & $<$-0.85 & -- & $<$-0.10 & $<$-0.36& --
 
\enddata
\tablecomments{68\% confidence widths are given when parameters are well determined, otherwise 3-$\sigma$ upper ($<$) or lower ($>$) value limits are listed.  No constraint (the probability spans the prior range) is denoted by  ``--" non-retrieved values are denoted by ``\nodata''. To aid interpretation, an uncertainty of 0.3 corresponds to a constraint within a factor of 2, 0.7 within a factor of 5, and 1 within a factor of 10 for the log value parameters. All values are uncertain to $\sim$10\% due to statistical jitter in MCMC retrievals and in the random noise instance applied to the simulated data. }
%\tablenotetext{a}{unconstrained lower abundances at 2-$\sigma$}
\tablenotetext{a}{Retrieved parameters for transmission (Trans) and emission (Emis) spectra are described in \S\ref{sec:Models}}
\tablenotetext{b}{68\% confidence width, but lower mixing ratio values are unconstrained at 2-$\sigma$}
\end{deluxetable*}

\begin{thebibliography}{}

\bibitem[Ag{\'u}ndez et 
al.(2014)]{APV14} Ag{\'u}ndez, M., Parmentier, V., Venot, O., Hersant, F., \& Selsis, F.\ 2014, \aap, 564, A73 

%\bibitem[Ag{\'u}ndez et al.(2014)]{AVS14} Ag{\'u}ndez, M., 
%Venot, O., Selsis, F., \& Iro, N.\ 2014, \apj, 781, 68 

\bibitem[Almenara et al.(2015)]{AAB15} Almenara, J.~M., 
Astudillo-Defru, N., Bonfils, X., et al.\ 2015, \aap, 581, L7

\bibitem[Barman et al.(2015)]{BKM15} Barman, T.~S., 
Konopacky, Q.~M., Macintosh, B., \& Marois, C.\ 2015, \apj, 804, 61 

\bibitem[Barstow et al.(2015)]{BAI15} Barstow, J.~K., Aigrain, S., Irwin, P.~G.~J., Kendrew, S., \& Fletcher, L.~N.\ 2015, \mnras, 451, 1306 

\bibitem[Barstow et al.(2013a)]{BAI13a} Barstow, J.~K., 
Aigrain, S., Irwin, P.~G.~J., et al.\ 2013, \mnras, 430, 1188 (2013a)

\bibitem[Barstow et al.(2013b)]{BAI13} Barstow, J.~K., 
Aigrain, S., Irwin, P.~G.~J., Fletcher, L.~N., 
\& Lee, J.-M.\ 2013, \mnras, 434, 2616 (2013b)

%\bibitem[Barstow et al.(2012)]{BTW12} Barstow, J.~K., Tsang, 
%C.~C.~C., Wilson, C.~F., et al.\ 2012, \icarus, 217, 542

%\bibitem[Batalha et al.(2014)]{BMK14} Batalha, N., Mandell,
%A., Kalirai, J., \& Clampin, M.\ 2014, Search for Life Beyond the Solar
%System.~Exoplanets, Biosignatures {\&} Instruments, P3P

\bibitem[Batalha et al.(2015)]{BKL15} Batalha, N., Kalirai, 
J., Lunine, J., Clampin, M., \& Lindler, D.\ 2015, arXiv:1507.02655 

\bibitem[Beichman et al.(2014)]{BBK14} Beichman, C., Benneke, 
B., Knutson, H., et al.\ 2014, \pasp, 126, 1134 

\bibitem[Benneke(2015)]{B15} Benneke, B.\ 2015, 
arXiv:1504.07655 

\bibitem[Benneke \& Seager(2012)]{BS12} Benneke, B., \& Seager, S.\ 2012, \apj, 753, 100 

\bibitem[Benneke \& Seager(2013)]{BS13} Benneke, B., \& Seager, S.\
2013, \apj, 778, 153

\bibitem[Brown(2001)]{B01} Brown, T.~M.\ 2001, \apj, 553, 1006

\bibitem[Burrows(2014)]{B14} Burrows, A.~S.\ 2014, 
Proceedings of the National Academy of Science, 111, 12601 

\bibitem[Burrows \& Orton(2010)]{BO10} Burrows, A., \&
Orton, G.\ 2010, Exoplanets, 419

\bibitem[Christiansen et al.(2010)]{CBC10} Christiansen, 
J.~L., Ballard, S., Charbonneau, D., et al.\ 2010, \apj, 710, 97

\bibitem[Cooper \& Showman(2006)]{CS06} Cooper, C.~S., \& Showman,
A.~P.\ 2006, \apj, 649, 1048

\bibitem[Cowan et al.(2015)]{CGA15} Cowan, N.~B., Greene, T., 
Angerhausen, D., et al.\ 2015, \pasp, 127, 311 

\bibitem[Croll et al.(2011)]{CAJ11} Croll, B., Albert, L., 
Jayawardhana, R., et al.\ 2011, \apj, 736, 78

\bibitem[Crossfield(2015)]{C15} Crossfield, I.~J.~M.\ 2015, 
\pasp, 127, 941 

\bibitem[Crossfield et al.(2015)]{CPS15} Crossfield, 
I.~J.~M., Petigura, E., Schlieder, J., et al.\ 2015, \apj, 804, 10 	

\bibitem[Crouzet et al.(2012)]{CMB12} Crouzet, N., 
McCullough, P.~R., Burke, C., \& Long, D.\ 2012, \apj, 761, 7

\bibitem[Deming et al.(2009)]{DSW09} Deming, D., Seager, S., 
Winn, J., et al.\ 2009, \pasp, 121, 952

\bibitem[Deming et al.(2013)]{DWM13} Deming, D., Wilkins, A., 
McCullough, P., et al.\ 2013, \apj, 774, 95

\bibitem[Diamond-Lowe et al.(2014)]{D-LSB14} Diamond-Lowe, H., 
Stevenson, K.~B., Bean, J.~L., Line, M.~R., 
\& Fortney, J.~J.\ 2014, \apj, 796, 66

\bibitem[Doyon et al.(2012)]{DHB12} Doyon, R., Hutchings, J.~B., Beaulieu, M.,
et al.\ 2012, \procspie, 8442

\bibitem[Ferruit et al.(2014)]{FBB14} Ferruit, P., Birkmann, 
S., B{\"o}ker, T., et al.\ 2014, \procspie, 9143, 91430A

\bibitem[Foreman-Mackey et al.(2014)]{F-MHM14} Foreman-Mackey, 
D., Hogg, D.~W., \& Morton, T.~D.\ 2014, \apj, 795, 64

\bibitem[Fortney(2005)]{F05} Fortney, J.~J.\ 2005, \mnras, 
364, 649

\bibitem[Fortney et al.(2011)]{FIN11} Fortney, J.~J., Ikoma, 
M., Nettelmann, N., Guillot, T., \& Marley, M.~S.\ 2011, \apj, 729, 32 

\bibitem[Fortney et al.(2008)]{FLM08} Fortney, J.~J., 
Lodders, K., Marley, M.~S., \& Freedman, R.~S.\ 2008, \apj, 678, 1419 

\bibitem[Fortney et al.(2013)]{FMN13} Fortney, J.~J., 
Mordasini, C., Nettelmann, N., et al.\ 2013, \apj, 775, 80 

\bibitem[Fraine et al.(2014)]{FDB14} Fraine, J., Deming, D., Benneke, B., et al.\ 2014, \nat, 513, 526 

\bibitem[Gibson et al.(2011)]{GPA11} Gibson, N.~P., Pont, F., 
\& Aigrain, S.\ 2011, \mnras, 411, 2199 

\bibitem[Gordon \& McBride(1994)]{GM94} Gordon, S., \& McBride, B.~J.\ 1994, NASA Reference Publ. 1311, http://www.grc.nasa.gov/WWW/CEAWeb/RP-1311.pdf

\bibitem[Greene et al.(2007)]{GBE07} Greene, T., Beichman, 
C., Eisenstein, D., et al.\ 2007, \procspie, 6693, 66930G 

\bibitem[Grimm \& Heng(2015)]{GH15} Grimm, S.~L., \&
Heng, K.\ 2015, \apj, 808, 182

\bibitem[Griffith(2014)]{G14} Griffith, C.~A.\ 2014, 
Philosophical Transactions of the Royal Society of London Series A, 372, 
30086

\bibitem[Han et al.(2014)]{HWW14} Han, E., Wang, S.~X., 
Wright, J.~T., et al.\ 2014, \pasp, 126, 827	

\bibitem[Hauschildt et al.(1999)]{HAB99} Hauschildt, P.~H., 
Allard, F., \& Baron, E.\ 1999, \apj, 512, 377

\bibitem[Haynes et al.(2015)]{HMM15} Haynes, K., Mandell, 
A.~M., Madhusudhan, N., Deming, D., \& Knutson, H.\ 2015, \apj, 806, 146 

\bibitem[Holland(1994)]{H94} Holland, H.~D.\ 1994, in Early Life on Earth, ed.
S. Bengtson, (New York, NY: Columbia University Press), 237

\bibitem[Heng \& Showman(2015)]{HS15} Heng, K., \& Showman,
A.~P.\ 2015, Annual Review of Earth and Planetary Sciences, 43, 509

\bibitem[Hu \& Seager(2014)]{HS14} Hu, R., \& Seager, S.\ 2014, \apj,
784, 63

\bibitem[Hu et al.(2013)]{HSB13} Hu, R., Seager, S., 
\& Bains, W.\ 2013, \apj, 769, 6 

\bibitem[Ida \& Lin(2005)]{IL05} Ida, S., \& Lin, D.~N.~C.\
2005, \apj, 626, 1045

\bibitem[Kataria et al.(2015)]{KSF15} Kataria, T., Showman, 
A.~P., Fortney, J.~J., et al.\ 2015, \apj, 801, 86

\bibitem[Kendrew et al.(2015)]{KSB15} Kendrew, S., Scheithauer, S.,
Bouchet, P., et al.\ 2015, \pasp, 127, 623 

\bibitem[Knutson et al.(2009)]{KCC09} Knutson, H.~A., 
Charbonneau, D., Cowan, N.~B., et al.\ 2009, \apj, 703, 769

\bibitem[Knutson et al.(2014)]{KDK14} Knutson, H.~A., 
Dragomir, D., Kreidberg, L., et al.\ 2014, \apj, 794, 155

\bibitem[Knutson et al.(2010)]{KHI10} Knutson, H.~A., Howard, 
A.~W., \& Isaacson, H.\ 2010, \apj, 720, 1569

\bibitem[Konopacky et al.(2013)]{KBM13} Konopacky, Q.~M., 
Barman, T.~S., Macintosh, B.~A., \& Marois, C.\ 2013, Science, 339, 1398

\bibitem[Kreidberg et al.(2014a)]{KBD14a} Kreidberg, L., Bean, 
J.~L., D{\'e}sert, J.-M., et al.\ 2014, \nat, 505, 69 (2014a)

\bibitem[Kreidberg et al.(2014b)]{KBD14b} Kreidberg, L., Bean, 
J.~L., D{\'e}sert, J.-M., et al.\ 2014, \apjl, 793, LL27 (2014b)

\bibitem[Kreidberg et al.(2015)]{KLB15} Kreidberg, L., Line, 
M.~R., Bean, J.~L., et al.\ 2015, arXiv:1504.05586

\bibitem[Lavvas et al.(2008)]{LCV08} Lavvas, P.~P., Coustenis, A., \&
Vardavas, I.~M.\ 2008, \planss, 56, 27

\bibitem[Lecavelier Des Etangs et al.(2008)]{DVD08} Lecavelier Des Etangs, A., Vidal-Madjar, A., D{\'e}sert, J.-M., \& Sing, D.\ 2008, \aap, 485, 865 

\bibitem[Lee et al.(2012)]{LFI12} Lee, J.-M., Fletcher, 
L.~N., \& Irwin, P.~G.~J.\ 2012, \mnras, 420, 170

\bibitem[Liang et al.(2003)]{LPL03} Liang, M.-C., Parkinson, 
C.~D., Lee, A.~Y.-T., Yung, Y.~L., \& Seager, S.\ 2003, \apjl, 596, L247 

\bibitem[Line et al.(2014b)]{LFM14} Line, M.~R., Fortney, 
J.~J., Marley, M.~S., \& Sorahana, S.\ 2014, \apj, 793, 33 (2014b)

\bibitem[Line et al.(2013b)]{LKD13} Line, M.~R., Knutson, H., 
Deming, D., Wilkins, A., \& Desert, J.-M.\ 2013, \apj, 778, 183 (2013b)

\bibitem[Line et al.(2014a)]{LKW14} Line, M.~R., Knutson, H., Wolf, A.~S., \& Yung, Y.~L.\ 2014, \apj, 783, 70 (2014a)

\bibitem[Line et al.(2010)]{LLY10} Line, M.~R., Liang, M.~C., 
\& Yung, Y.~L.\ 2010, \apj, 717, 496 

\bibitem[Line et al.(2015)]{LTB15} Line, M.~R., Teske, J., 
Burningham, B., Fortney, J.~J., \& Marley, M.~S.\ 2015, \apj, 807, 183 

\bibitem[Line et al.(2011)]{LVC11} Line, M.~R., Vasisht, G., 
Chen, P., Angerhausen, D., \& Yung, Y.~L.\ 2011, \apj, 738, 32 

\bibitem[Line et al.(2013a)]{LWZ13} Line, M.~R., Wolf, A.~S., 
Zhang, X., et al.\ 2013, \apj, 775, 137 (2013a)

\bibitem[Line \& Yung(2013)]{LY13} Line, M.~R., \& Yung, Y.~L.\ 2013, \apj, 779, 3

\bibitem[Line et al.(2012)]{LZV12} Line, M.~R., Zhang, X., 
Vasisht, G., et al.\ 2012, \apj, 749, 93 

\bibitem[Madhusudhan(2012)]{M12} Madhusudhan, N.\ 2012, 
\apj, 758, 36 

\bibitem[Madhusudhan et al.(2011)]{MHS11} Madhusudhan, N., 
Harrington, J., Stevenson, K.~B., et al.\ 2011, \nat, 469, 64 

\bibitem[Madhusudhan et al.(2014)]{MKF14} Madhusudhan, N.,
Knutson, H., Fortney, J.~J., \& Barman, T.\ 2014, Protostars and Planets
VI, 739

\bibitem[Madhusudhan \& Seager(2009)]{MS09} Madhusudhan, N., \& Seager,
S.\ 2009, \apj, 707, 24

%\bibitem[Madhusudhan \& Seager(2010)]{MS10} Madhusudhan, N., \& Seager, S.\ 2010, \apj, 725, 261 

\bibitem[McCullough et al.(2014)]{MCD14} McCullough, P.~R., 
Crouzet, N., Deming, D., \& Madhusudhan, N.\ 2014, \apj, 791, 55 

\bibitem[Miller \& Fortney(2011)]{MF11} Miller, N., \& Fortney, J.~J.\ 2011,
\apjl, 736, L29

\bibitem[Miller-Ricci Kempton et al.(2012)]{MZF12} Miller-Ricci Kempton,
E., Zahnle, K., \& Fortney, J.~J.\ 2012, \apj, 745, 3

\bibitem[Mordasini et al.(2012)]{MAK12} Mordasini, C., Alibert, Y., Klahr, H.,
\& Henning, T.\ 2012, \aap, 547, A111

\bibitem[Moses et al.(2013)]{MLV13} Moses, J.~I., Line, 
M.~R., Visscher, C., et al.\ 2013, \apj, 777, 34 

\bibitem[Moses et al.(2011)]{MVF11} Moses, J.~I., Visscher, 
C., Fortney, J.~J., et al.\ 2011, \apj, 737, 15 

\bibitem[{\"O}berg et al.(2011)]{OMB11} {\"O}berg, K.~I., 
Murray-Clay, R., \& Bergin, E.~A.\ 2011, \apjl, 743, L16

\bibitem[Pont et al.(2013)]{PSG13} Pont, F., Sing, D.~K., 
Gibson, N.~P., et al.\ 2013, \mnras, 432, 2917

\bibitem[Prinn \& Barshay(1977)]{PB77} Prinn, R.~G., \&
Barshay, S.~S.\ 1977, Science, 198, 1031

\bibitem[Schwieterman et al.(2015)]{SRM15} Schwieterman, E.~W.,
Robinson, T.~D., Meadows, V.~S., Misra, A., \& Domagal-Goldman, S.\
2015, \apj, 810, 57

\bibitem[Seager \& Deming(2010)]{SD10} Seager, S., \&
Deming, D.\ 2010, \araa, 48, 631

\bibitem[Schwarz et al.(2015)]{SBd15} Schwarz, H., Brogi, M., de Kok,
R., Birkby, J., \& Snellen, I.\ 2015, \aap, 576, A111

\bibitem[Shabram et al.(2011)]{SFG11} Shabram, M., Fortney, 
J.~J., Greene, T.~P., \& Freedman, R.~S.\ 2011, \apj, 727, 65 

\bibitem[Stevenson et al.(2014)]{SBM14} Stevenson, K.~B., 
Bean, J.~L., Madhusudhan, N., \& Harrington, J.\ 2014, \apj, 791, 36

\bibitem[Stevenson et al.(2010)]{SHN10} Stevenson, K.~B., 
Harrington, J., Nymeyer, S., et al.\ 2010, \nat, 464, 1161

\bibitem[Swain et al.(2014)]{SLD14} Swain, M.~R., Line, 
M.~R., \& Deroo, P.\ 2014, \apj, 784, 133

%\bibitem[Swain et al.(2009)]{STV09} Swain, M.~R., Tinetti, 
%G., Vasisht, G., et al.\ 2009, \apj, 704, 1616

\bibitem[Swain et al.(2009)]{SVT09} Swain, M.~R., Vasisht, 
G., Tinetti, G., et al.\ 2009, \apjl, 690, L114 

\bibitem[Venot et al.(2014)]{VAS14} Venot, O., Ag{\'u}ndez, M., Selsis,
F., Tessenyi, M., \& Iro, N.\ 2014, \aap, 562, A51

\bibitem[Venot et al.(2012)]{VHA12} Venot, O., H{\'e}brard,
E., Ag{\'u}ndez, M., et al.\ 2012, \aap, 546, A43

\bibitem[Visscher \& Moses(2011)]{VM11} Visscher, C., \& Moses, J.~I.\ 2011, \apj, 738, 72

\bibitem[Waldmann et al.(2015)]{WTR15} Waldmann, I.~P., 
Tinetti, G., Rocchetto, M., et al.\ 2015, \apj, 802, 107

\bibitem[Yung \& Demore(1999)]{YD99} Yung, Y.~L., \& Demore, W.~B.\ 1999,
Photochemistry of planetary atmospheres / Yuk L.~Yung, William
B.~DeMore.~New York : Oxford University Press, 1999.~ QB603.A85 Y86 1999

\bibitem[Zahnle et al.(2009)]{ZMF09} Zahnle, K., Marley, M.~S., Freedman,
R.~S., Lodders, K., \& Fortney, J.~J.\ 2009, \apjl, 701, L20

\end{thebibliography}
\end{document}